\documentclass[draft]{agujournal2019}
\usepackage{url} 
\usepackage{lineno}
\usepackage[inline]{trackchanges} 
\usepackage{soul}
\usepackage{amsmath}
\usepackage{amsfonts}
\usepackage{multirow}
\usepackage{multicol}
\draftfalse

\journalname{JGR: Machine Learning and Computation}

\begin{document}

%
%


\title{Using Explainable AI and Transfer Learning to understand and predict the maintenance of Atlantic blocking with limited observational data}

%
%




\authors{Huan Zhang\affil{1}, Justin Finkel\affil{2}, Dorian S. Abbot\affil{3},  Edwin P. Gerber\affil{1} , and Jonathan Weare\affil{1}   }

\affiliation{1}{Courant Institute of Mathematical Sciences, New York University}
\affiliation{2}{Department of Earth, Atmospheric, and Planetary Sciences, Massachusetts Institute of Technology}
\affiliation{3}{Department of the Geophysical Sciences, University of Chicago}




\correspondingauthor{Jonathan Weare}{weare@nyu.edu}



\begin{keypoints}
     \item Given sufficient training data, convolutional neural networks can predict the maintenance of Atlantic blocking from an initial blocked state.
     \item Transfer learning from an idealized model to reanalysis data enables predictive skill in the low data regime of the observational record.
    \item Feature importance analysis reveals the influence of upstream flow on blocking persistence and quantifies biases in the idealized model.

\end{keypoints}

%
%

%
%


\begin{abstract}
Blocking events are an important cause of extreme weather, especially long-lasting blocking events that trap weather systems in place. The duration of blocking events is, however, underestimated in climate models. Explainable Artificial Intelligence are a class of data analysis methods that can help identify physical causes of prolonged blocking events and diagnose model deficiencies. We demonstrate this approach on an idealized quasigeostrophic model developed by Marshall and Molteni (1993). We train a convolutional neural network (CNN), and subsequently, build a sparse predictive model for the persistence of Atlantic blocking, conditioned on an initial high-pressure anomaly. Shapley Additive ExPlanation (SHAP) analysis reveals that high-pressure anomalies in the American Southeast and North Atlantic, separated by a trough over Atlantic Canada, contribute significantly to prediction of sustained blocking events in the Atlantic region. This agrees with previous work that identified precursors in the same regions via wave train analysis. When we apply the same CNN to blockings in the ERA5 atmospheric reanalysis, there is insufficient data to accurately predict persistent blocks. We partially overcome this limitation by pre-training the CNN on the plentiful data of the Marshall-Molteni model, and then using Transfer Learning to achieve better predictions than direct training. SHAP analysis before and after transfer learning allows a comparison between the predictive features in the reanalysis and the quasigeostrophic model, quantifying dynamical biases in the idealized model. This work demonstrates the potential for machine learning methods to extract meaningful precursors of extreme weather events and achieve better prediction using limited observational data.
\end{abstract}

\section*{Plain Language Summary}
Blocking events are an important cause of extreme weather, especially long-lasting blocking events that trap weather systems in place. The duration of blocking events is, however, systematically underestimated in climate models.  Using data generated by a simplified atmospheric model we demonstrate that, given sufficient training data, convolutional neural networks can predict the maintenance of Atlantic blocking from an initial blocked state. Next, we show that first training the neural network on data from the simplified model and then fine tuning the training using real world weather data enables prediction even with few examples of long-lasting blocking events in the observational record.
 Subsequent feature analysis of the resulting neural networks identifies the input variables that most strongly impact their predictions, revealing that areas of high pressure in certain parts of North America and the North Atlantic Ocean are important for predicting long-lasting blocking events and quantifying biases in the idealized model relative to real weather.
\section{Introduction}
Blocking events are high-amplitude, quasi-stationary anticyclonic high-pressure anomalies that give rise to prolonged abnormal weather conditions in the mid-to-high latitudes~\cite{Rex1950,Woollings2018, Lupo2021}. Blocking events can lead to regional extreme weather by disrupting the usual westerly flow for extended periods~\cite<e.g.,>[]{Woollings2022}, causing extreme heatwaves, floods, and  winter storms~\cite<e.g.,>[]{lupo2012dynamic}.

The predictive skill of numerical weather models has improved dramatically, but they still cannot accurately forecast important aspects of blocking events. Blocking frequency and duration are generally simulated poorly by climate models~\cite{ClimateDavini2020}, and even by numerical weather prediction models in medium-range forecasts~\cite{Woollings2018,  Mediumrange2009, Ferranti2015blforcast}.  Several possible contributing factors have been proposed, including the accuracy of the model's mean flow~\cite{scaife2010atmospheric} or synoptic eddies~\cite{berckmans2013atmospheric, zappa2014linking},  the model's resolution~\cite{davini2016northern} and  subgrid-scale parameterizations~\cite{d1998northern}, and even the choice of blocking index itself~\cite{tibaldi1990operational,dole1983persistent,pelly2003new}.

Two commonly used blocking indices~\cite{tibaldi1990operational,dole1983persistent} highlight two essential features of a blocking \emph{event}: (i) a large positive anomaly of geopotential height that displaces the midlatitude jet, ``blocking" the flow, that (ii) persists for longer than typical synoptic variability.  Often a 5 day threshold is invoked, but the longer the flow remains in a blocked state, the more severe the implications, either for extended cold/hot conditions or an increased likelihood of compound storm events \cite<e.g., back-to-back storms, which can dramatically increase the potential for damage;>[]{Woollings2022}.   The persistence of blocking is the focus of our study: given the onset of a blocked state, what is the likelihood that the flow will remain blocked for an extended period, 5 days for a standard event, or up to 9 days for more extreme cases?  We take a data-driven approach, training a convolutional neural network to identify persistent blocks at the onset of a blocked state.


To understand blocking, various low-order models have been formulated to identify essential features. In an influential early work, \citeA{charney1979multiple} modeled blocking as one of two equilibrium states of a set of dynamical equations for a highly truncated barotropic channel model. Others used low-order models to propose that the positive feedback of synoptic-scale eddies on the blocking structure contributes to the long-time maintenance of blocks~\cite{Hoskins1983,shutts1983propagation,MCWILLIAMS198043}. While these low-order models have provided useful physical insight, realistic land-sea interactions, topography, and other factors present in the real world limit their application. Comprehensive models, on the other hand, are becoming skillful in simulating realistic blocking, but their complexity makes it challenging to isolate the essential mechanism(s), and expensive to simulate numerous events. 

To strike a balance between complexity, transparency, and statistical robustness from abundant data (model output), we begin with the  Marshall-Molteni (MM) model~\cite{marshall1993toward}, a three-layer quasigeostrophic (QG) approximation of the atmosphere that has previously been used to study blocking events~\cite<e.g.,>[]{lucarini2020new}. The MM model captures the main features of the northern hemisphere atmosphere reasonably well. For example, \citeA{michelangeli1998dynamics} found that an enhanced baroclinic wavetrain traveling across the North Atlantic is necessary to trigger the onset of the Euro-Atlantic blocking in both this simple model and reanalysis. They also pointed out that wave-wave interactions and wave-mean interactions dominate local amplification and the propagation of anomalies, respectively.


The MM model allows us the freedom to develop and test methods in a data-rich setting.  How well can a data-driven method identify persistent events as a function of the input data you allow it?  Following work by \citeA{explanableAIBarnes} and \citeA{RAMPAL2022100525}, can so-called Explainable Artificial Intelligence (XAI) techniques provide  physical insight into both the AI methods and the model itself? We show that Shapley Additive ExPlanation (SHAP) analysis reveals key regions upstream of the blocking center that enable prediction, and use this to construct low-order models the can be interpreted in the context of prior work.

Our ultimate goal, however, is to forecast and understand the maintenance of blocks in our atmosphere, for which we shift the focus to ERA5 reanalysis~\cite{ERA5reanalysis2020}.  For the most extreme case of a 9-day block in the North Atlantic, only 18 have occurred in the historical record (See Tab.~\ref{tab:ERA5dataset}). What chance does a data-driven approach have?  To address the problem of limited data, we apply transfer learning: first we train a convolutional neural network on the MM model to learn the basic features of blocking, and then we re-train it on the limited ERA5 data to calibrate it for the real atmosphere.  We find that pre-training on the MM model yields a better predictor than when we train the same network on ERA5 alone, proving the efficacy of the transfer learning approach.


The remainder of this paper is organized as follows. Section~\ref{sec:MMmodel} introduces the Marshall-Molteni (MM) model. Sections ~\ref{sec:BI} and~\ref{sec:eventdef} define our choice of blocking index and blocking event criteria, and formulate an objective function for machine learning. Section~\ref{sec:cnn} discusses our convolutional neural network structure and training details. We first focus exclusively on the MM model in sections~\ref{sec:XAI} and ~\ref{sec:sparsemodel}, applying XAI techniques to visualize the important features for prediction and testing the results by building a sparse model with features guided by the XAI. We also suggest physical interpretations for these predictive features. Finally, we turn to the ERA5 data set in Section~\ref{sec:transferlearning}, applying transfer learning to improve the prediction of persistent blocks in ERA5, especially for more extreme events.  SHAP analysis shows how transfer learning has modified the CNN to adapt to the new data set, but preserves the use of key upstream regions for prediction.

\section{Marshall-Molteni Model}
\label{sec:MMmodel}
~\citeA{marshall1993toward} developed a 3-layer model of the atmosphere to study  atmospheric low-frequency variability. We use a Northern Hemisphere only version of the model developed by~\citeA{lucarini2020new} with 6210 degrees of freedom. We refer the reader to that paper for a complete description, but review key details here.
The Marshall-Molteni (MM) model state is specified by potential vorticity $q_j$ in three layers of the atmosphere, $j=1,2,3$, corresponding to pressure levels 200, 500, and 800 hPa. $q_j$ evolves according to quasi-geostrophic dynamics as
\begin{align}
    \partial_t q_j+J(\psi_j, q_j)=-D_j+S_j
\end{align}
where $\psi_j$ is the streamfunction in layer $j$, related to $q_j$ as
\begin{align}
    q_1&=\Delta \psi_1-(\psi_1-\psi_2)/R_1^2 +f\\
    q_2&=\Delta \psi_2+(\psi_1-\psi_2)/R_1^2-(\psi_2-\psi_3)/R_2^2 +f\\
    q_3&=\Delta \psi_3+(\psi_2-\psi_3)/R_2^2 +f(1+h/H_0).
\end{align}
Here, $\Delta$ is the horizontal Laplacian operator, $R_1=761$ km and $R_2=488$ km are the Rossby deformation radii in layers 1 and 2, $f=2\Omega \cos \phi$ is the latitude-dependent Coriolis parameter, and $h$ is the orography of the surface, rescaled by the constant $H_0$. The operator $D_j$ combines all dissipative terms, including radiative damping, surface friction and hyper-diffusion to crudely parametrize small scale diffusion, but is also necessary for numerical stability:
\begin{align}
    \begin{split}
        -D_1=&(\psi_1-\psi_2)/(\tau_R R_1^2)-R^8\Delta^4q_1/(\tau_H \lambda_{max}^4)\\
        -D_2=&-(\psi_1-\psi_2)/(\tau_R R_1^2)+(\psi_2-\psi_3)/(\tau_R R_2^2) -R^8 \Delta^4 q_2'/(\tau_H \lambda^4_{max})\\
        -D_3=&-(\psi_2-\psi_3)/(\tau_R R_2^2)-EK_3-R^8 \Delta^4 q_3'/(\tau_H\lambda_{max}^4).
    \end{split}
\end{align}
The forcing, $S_j$ is computed from observed data to inject energy into the system and give the model a realistic mean state:
\begin{align}
    S_j=\overline{J(\psi_j, q_j)}+\overline{D}_j
\end{align}

The data to construct $S_j$ were drawn from the 1983–1992 winter (DJF) climatology of the ERA40 reanalysis provided by ECMWF. The model is run with T31 horizontal resolution (corresponding to 90 longitude $\times$ 23 latitude gridpoints across the northern hemisphere).
All model output fields, as well as the reanalysis used later, are averaged daily.  The climatology of Marshal-Molteni model is shown in the supplemental materials, and we compare its blocking statistics with ERA5 reanalysis in the next section.

\section{Blocking index}
\label{sec:BI}

In this study, we use the ``DG" index~\cite{dole1983persistent} to define blocking events. This is an anomaly-based blocking index, but has been shown to capture the same essential features of blocking as other indices, e.g., that of~\citeA{tibaldi1990operational}.

We compute this index by transforming the spherical harmonic representation of $\psi$ into approximate geopotential height, $Z$, on a Gaussian grid for latitude and a uniform grid for longitude. The approximation is the choice of a fixed Coriolis parameter $f_0$ to convert from $\psi$ to $Z$, which leads to minimal distortion over our midlatitude area of focus. A blocking event is said to occur at a specific location when $Z$ stays above a tunable geopotential height anomaly threshold, $M$, for at least 5 consecutive days. In their paper, \citeA{dole1983persistent} tested statistics for varying $M$ values, ranging from 50 m to 250 m, with subsequent studies adopting different thresholds~(\citeA{Pedramblockingindices}, Tab. 2). For our investigation, we calibrated $M=100$ m for our MM model simulation to roughly match the blocking fraction computed from ERA5 reanalysis data, where we used the threshold $M=150$ m as in \citeA{mullen1987transient}.

\begin{figure}
    \centering
   \includegraphics[width=0.9\columnwidth]{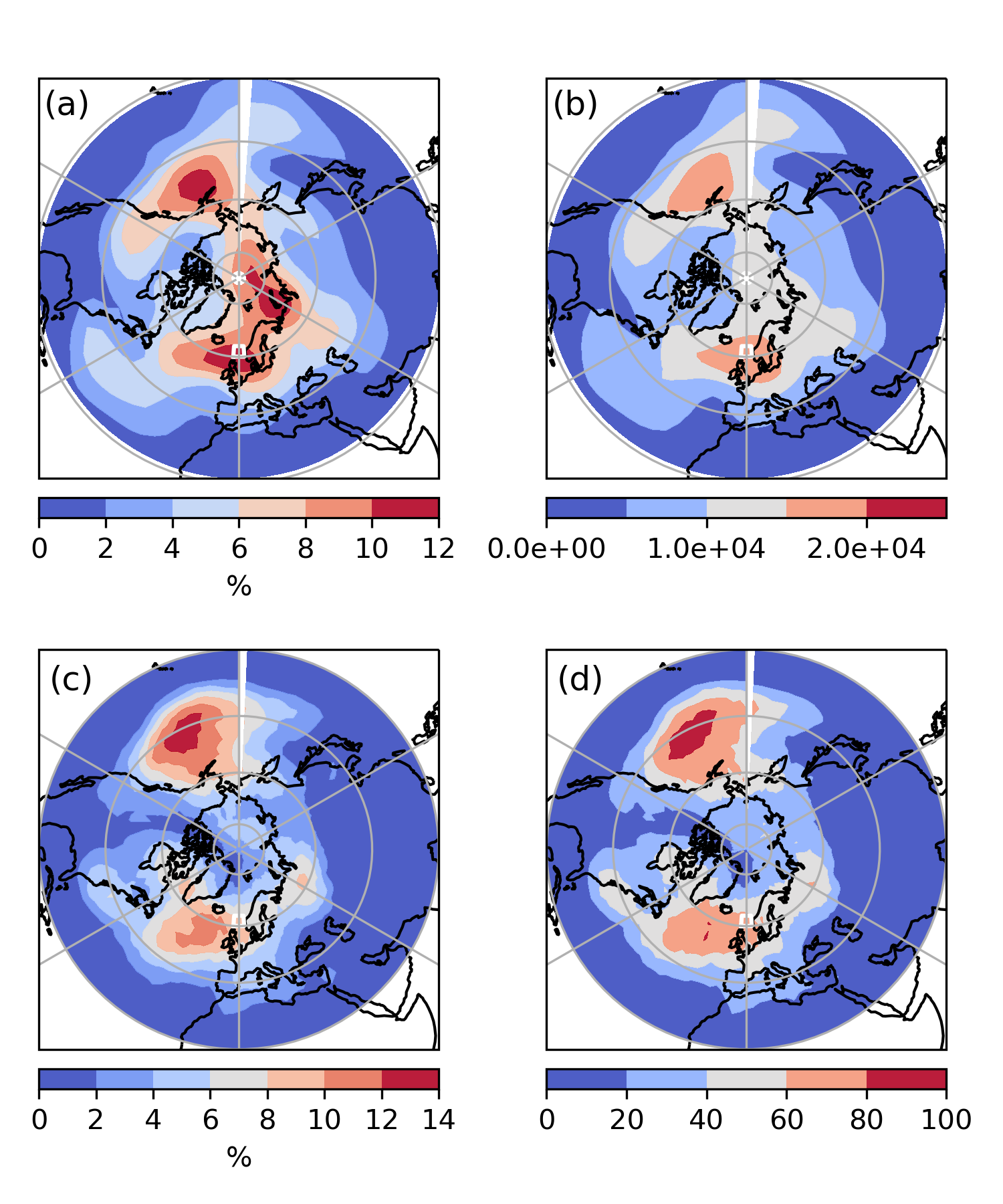}
   \caption{ (a) blocking fraction (the percent of days with $T\geq5$ days) for MM model data with $M=100$ m. (b) total blocking event counts for MM model data during the simulation. (c) blocking fraction for ERA5 reanalysis data with $M=150$. (d) total blocking event for ERA5 reanalysis data with $M=150$m. In all  subfigures, the  region we focus on is indicated by the white rectangle centered at $0^\circ$E and $62^\circ N$ (approximately spanned by 3 longitude points covering $4^\circ W-4^\circ E$, and 2 latitude points covering $60^\circ N-64^\circ N$)}
   \label{fig:Fig1:blocking_statistics}
\end{figure}

Fig.~\ref{fig:Fig1:blocking_statistics} shows the blocking event statistics during the simulation.  For comparison, blocking event statistics computed from ERA5 reanalysis data from 1959-2021 are also shown. 
In this study, we focus on North Atlantic blockings indicated by the white rectangle in Fig.~\ref{fig:Fig1:blocking_statistics}. We pick this region because it has a relatively high blocking frequency, and for its important influence on western Europe. 
We use $Z_{B}$, the mean 500 hPa geopotential height anomaly in this target region over the North Atlantic, to define blocked states and blocking events.

\section{Probabilistic forecasting and event definition}
\label{sec:eventdef}
\begin{figure}
    \centering
   \includegraphics[width=1.0\columnwidth]{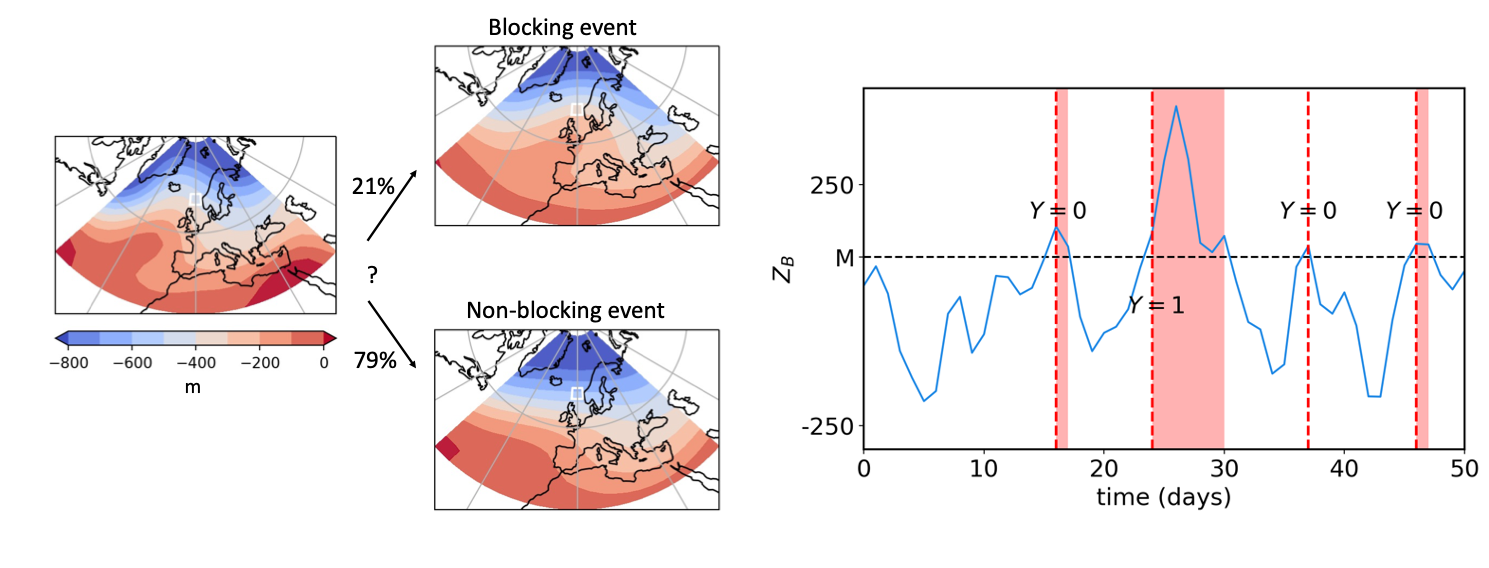}
   \caption{Left: The blocking persistence problem: given a nascent blocked state, the goal is to forecast whether it will persist into a long-lasting blocking event, or quickly return to climatology. The percentile represents the climatological probability. Right: A sample trajectory of $Z_B(t)$, the anomaly of geopotential height defined in Sec.~\ref{sec:BI}. The vertical dashed lines indicate new blocked states ($T=1$). The red shading indicates the duration of the block.  The label $Y=1$ indicates that the blocked state persisted 5 days to constitute a blocking event, while $Y=0$ indicates that it did not.}
   \label{fig:Fig2:forecast}
\end{figure}

We aim to study the \emph{maintenance} of blocks rather than their \emph{onset}. Precisely, we formulate the question as the classification problem posed in Fig.~\ref{fig:Fig2:forecast}: given a nascent blocked state, i.e., the state on a day that geopotential height anomalies over the North Atlantic first exceed the threshold $M$, can we immediately predict whether the flow will remained blocked for 5 or more days -- evolving into a blocking \emph{event} -- or will the flow return back towards the climatological state before 5 days have passed?  In the MM model, nascent blocked states evolve into 5-day persistent blocking events approximately 1/5th (21\%) of the time on average, more often fading back towards climatology.  Given only the state at the time of blocking onset, can a data-driven method accurately identify the rarer cases that will persist for more than 5 consecutive days?  

To formulate this classification problem mathematically, we denote the full model state by $\boldsymbol{X}$ and further introduce a variable $T$ for the running duration of a blocked state:
\begin{align}
    T=\{\text{days since }  Z_B<M\}.
    \label{eq:defT}
\end{align}
Note that $Z_B(t)$ is determined by the state vector $\boldsymbol{X}(t)$ at any time $t$, but $T(t)$ retains some memory of previous states and thus is not fully determined by $\boldsymbol{X}(t)$.
For example, as shown in Fig.~\ref{fig:Fig2:forecast}, suppose $Z_B(t)$ first rises above $M$ on day $t=16$ and dips back below $M$ on day $t=18$. Then, $T(t)=0$ for all days through $t=15$, $T(16)=1$, $T(17)=2$, and $T(18)=0$.
With this notation, we can say that ``$\boldsymbol{X}(t)$ is the beginning of a blocking event'' if
\begin{align}\label{def:startsblocking}
    T(t)=1\quad \text{and}\quad T(t+D-1) =D.
\end{align}
The condition $T(t+D-1) =D$ only holds when there are at least $D$ consecutive  days with $Z_B(t)\geq M$ starting from $t$. We can see an example of this in Fig.~\ref{fig:Fig2:forecast} at day 24, for both a block of duration 5 and 7 days.  Here, $T(24)=1$, and $T(28)=5$, triggering the condition for $D=5$.  The flow remains blocked through $T(30)=7$, such that day 24 would also count as the onset of a $D=7$ day blocking event.   

With this formulation, our central question becomes: given a $T(t)=1$ state at time $t$ (the flow has just become blocked), will it stay blocked for $D$ days, $T(t+D-1) =D$, or not? We address this question by estimating the
conditional probability:
\begin{equation}\label{eq:q}
    q(\boldsymbol{x}(t))=\mathbb{P}[T(t+D-1) =D\,|\,\boldsymbol{X}(t)=\boldsymbol{x}(t),T(t)=1].
\end{equation}
Unless otherwise specified, we adopt $D=5$ to maintain consistency with the common blocking indices~\cite{tibaldi1990operational,dole1983persistent,pelly2003new}. We also consider more extreme events with $D=7$ and $D=9$.

\section{Convolutional Neural Network Training and Performance}
\label{sec:cnn}
Convolutional Neural Networks (CNN) have gained widespread application in probabilistic forecasting problems~\cite{Bouchet2023,Ham2019,liu2016application} for their outstanding performance on multidimensional data sets with spatial structure.  A CNN differs from a dense neural network in the use of convolutional layers with shared weights and biases across layers within the network, designed to extract features that exhibit translation invariance across the input space~\cite{Goodfellow-et-al-2016}.  Originally developed in the context of image processing, CNN excels in scenarios where target features, such as the face of a cat, may appear at different places within the training image.  Convolutional layers allow the network to efficiently learn these features, combining information across multiple images.  In our context, atmospheric eddies and Rossby waves share similar dynamics across all longitudes.  A CNN can potentially more effectively extract these dynamics, while still learning how they vary with longitude and zonal asymmtries induced by topography, etc.

\begin{figure}
    \centering
   \includegraphics[width=0.5\columnwidth]{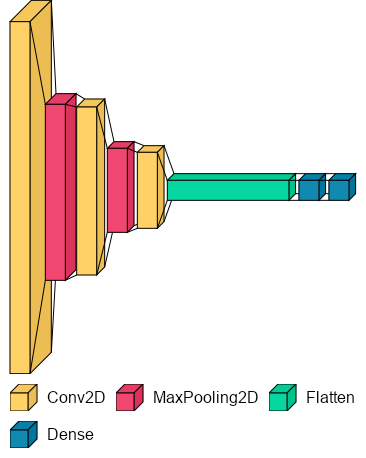}
   \caption{ Convolutional Neural Network structure. The three convolutional layers (yellow) respectively use 32, 64 and 64 filters ($3\times 3$), followed by ReLu activations. Between each pair of convolutional layers is a max-pooling layer (red) with window size $2\times 2$. Then the output is flattened (green) and passed to a dense layer with 64 neurons that produces 2 outputs (blue). The output is then passed through a softmax function (blue).}
   \label{fig:CNN structure}
\end{figure}

The structure of the CNN in this investigation follows \citeA{Bouchet2023} and is shown in Fig.~\ref{fig:CNN structure}. It consists of a three-layer architecture, combining convolutional filters followed by ReLu activations. Specifically, we use 32 and 64 filters ($3\times 3$)  for the first and last two convolutional layers. Between each pair of convolutional layers is a max-pooling layer. The output is then flattened and passed to a dense layer with 64 neurons that produces 2 outputs. The output is then passed through a softmax function to form two normalized probabilities that sum to 1.

We performed experiments with alternative CNN structures and found that reducing the widths of layers can mitigate overfitting, but this also reduces the performance at the best epoch (not shown). Therefore we adopt the architecture in Fig.~\ref{fig:CNN structure}  and use early-stopping  to avoid overfitting, as detailed below. 

\subsection{Training and Test Datasets}

To study whether a nascent blocked state will persist, we create a training and test set of all states  where the flow has just become blocked: $\{(\boldsymbol{X},T)|T=1\}$, where $\boldsymbol{X}$ are $18\times 90\times 3$ (latitudes $\times$ longitudes $\times$ pressure at levels of 200 hPa, 500 hPa, 800 hPa) grid maps of geopotential height from 20$^\circ $N to 87$^\circ $N.
Our goal is to classify which of these cases  persist into blocking events ($Y=1$) versus states that do not ($Y=0$). Fig.~\ref{fig:Fig2:forecast} shows a sample time series with 4 instances of a nascent blocked state, $t=16, 24, 38$ and 47, only the second of which evolves into a persistent blocking event, $Y=0$, 1, 0, and 0, respectively. For each case, the model must classify $Y=0$ or $Y=1$ given only $\boldsymbol{X}$ at the onset time.

We examined the sensitivity of CNN model performance with respect to different amounts of training data. To prepare the dataset, we integrate the MM model for 1250k days in total. The computational cost is low, requiring 1 CPU core and approximately 11 hours. We select the first $n$ days (with $n$ ranging from 1k to 1000k) to create the training data set, and always take the last 250k days for the test dataset. Thus all models can be fairly compared. The trajectory length and the corresponding number of nascent blocked state states are shown in Tab.~\ref{tab:traj_vs_numberofdata}. The likelihood $q$ of forming a blocking event varies depending on different persistence thresholds $D$. This dependence relationship is illustrated in Tab.~\ref{tab:MMdataset}.
\begin{table}
    \centering
    \begin{tabular}{cc} \hline
        \multicolumn{2}{c}{Training data}\\ \hline
         Days & Nascent blocked states \\ \hline
         1k& 63 \\
         10k& 699 \\
         100k& 7024 \\
         500k& 35078 \\
         1000k& 70635 \\ \hline
    \end{tabular}
    \quad
    \begin{tabular}{cc} \hline
        \multicolumn{2}{c}{Test data}\\ \hline
         Days & Nascent blocked states \\ \hline
           &   \\
            &   \\
           250k   &17755\\
           &   \\
            &   \\\hline
    \end{tabular}
    \caption{Length of trajectory (in thousands of days) vs.~number of nascent blocking states ($T=1$) in training set and test sets of varying size.}
    \label{tab:traj_vs_numberofdata}
\end{table}

\subsection{Learning procedure}\label{sub:learningprocedure}

For simplicity, we use binary cross entropy as a loss function, a common choice for classification~\cite{Bouchet2023}. Alternative loss functions have been studied by~\citeA{rudy2023output}. The loss function $L(q)$ is defined as as follows:
\begin{align}
    L(q)=-\frac{1}{N}\sum_{i=1}^N&\Big[Y_i \log q(Y_i=1)+(1-Y_i) \log (1-q(Y_i=1))\Big] \nonumber
\end{align}
where $q(Y_i=1)\in (0,1)$
is the probability of the event $Y_i=1$ as predicted by the CNN. $L(q)$ is small when the CNN predicts high probability for positive events, and low probability for negative events.

Given the rarity of blocking events, the data exhibit a pronounced class-imbalance, which becomes increasingly severe for longer block durations. As shown in Tab.~\ref{tab:MMdataset}, for $D=5$, only about 1 in 5 nascent blocked states persist into an event, but $D=9$, less than 1 in 20 evolve into persistent events. With this extreme imbalance, a model that never predicts an event will be correct over 80\% or 95\% of the time, respectively. However, such a model would clearly underperform in terms of precision and recall, which would both be zero.
\begin{table}
\centering
\begin{tabular}{cccc}  \hline
         Threshold&$Y=1$& $Y=0$ &Positive rate \\
         \hline
         $\geq 5$ days&  18748&  69642 &0.212 \\
         $\geq 7$ days&8522&79868 &0.096 \\
        $\geq 9$ days& 3891&84499 &0.044 \\
 \hline
\end{tabular}
    \caption{The statistics of blocking events in our MM 1250k day simulation.  The full dataset exhibits 88390 nascent blocking states ($T=1$ states).  $Y=1$ marks the number of these nascent blocks that persist for 5, 7, or 9 days, thus evolving into a blocking event under these respective thresholds, while $Y=0$ denotes the number that don't make it to the threshold. }
    \label{tab:MMdataset}
\end{table}

To address the class imbalance, for our results in this section we employ over-sampling~\cite{johnson2019survey} techniques during training. In each epoch, we sample an equal number of nascent blocks from both classes until we complete an iteration over all the nascent blocks in the overrepresented class. As a result, the nascent blocks that persist have been sampled multiple times during each epoch.

\subsection{Performance metrics}
Throughout this study, we evaluate model performance using two key metrics: \emph{precision} and \emph{recall}. We monitor the values of these metrics on the test dataset throughout the training process to determine the stopping point in order to avoid overfitting.
The precision and recall are respectively defined as
\begin{alignat}{2}
    &\text{Precision}&&=\frac{\text{True positives}}{\text{True positives + False positives}},\\
    &\text{Recall}&&=\frac{\text{True positives}}{\text{True positives + False negatives }},
\end{alignat}
where ``True positives" is the number of data points with $Y=1$ for which our CNN predicts a persistent blocking event, ``False positives" the number of data points with $Y=0$ for which our CNN predicts a persistent blocking event, and ``False negatives" the number of data points with $Y=1$ for which our CNN predicts a  blocked state that does not persist.

More informally, if the method forecasts that an event will occur, the precision measures the fraction of times this forecast is correct.  The recall, on the other hand, is the fraction of the successfully forecasted events of all the positive events.
If, regardless of the system state, one randomly predicts events with the climatological mean rate, in which an overall fraction $p$ of the data labels are True, then the precision and recall are both given by $\frac{p^2N}{p^2N+(1-p)pN}=p$. This sets the floor for a useful predictor: both the precision and recall must be better than climatological rate. 

There can be tradeoffs between improving the precision and recall. Predicting the event all the time will give you a perfect recall, but climatological precision $p$. A low recall implies missing a substantial number of positive events, leading to inadequate preparation and increased risk of damage. Conversely, a low precision suggests over-predicting events, ``crying wolf" too often. In the context of extreme weather forecasting, this can lead to over-preparation, consequently reducing the efficiency of regular societal operations, as well as trust.

A reasonably high value of both recall and precision is crucial for an effective and resource-efficient forecasting model. We use a simplistic definition of `best' performance, expressed as
\begin{align}
\label{eq:bestperformance}
    \text{Overall performance}=\text{Precision}+\text{Recall}\,.
\end{align}
However, it is crucial to note that in practical scenarios, designing overall performance metrics requires careful consideration of the cost of preparing vs.~risk of damage associated without preparation. 
This naive criteria only works when the precision and recall are both reasonably high, since forecasting the event all the time will yield a performance score of 1+p (recall of 1 and precision of $p$). We used caution in ERA5 based forecasts, requiring our trained models exhibit nontrivial precision above the climatological rate.

\subsection{Performance and early stopping technique}
\label{sec:earlystopping}
\begin{figure}
    \centering
   \includegraphics[width=1.0\columnwidth]{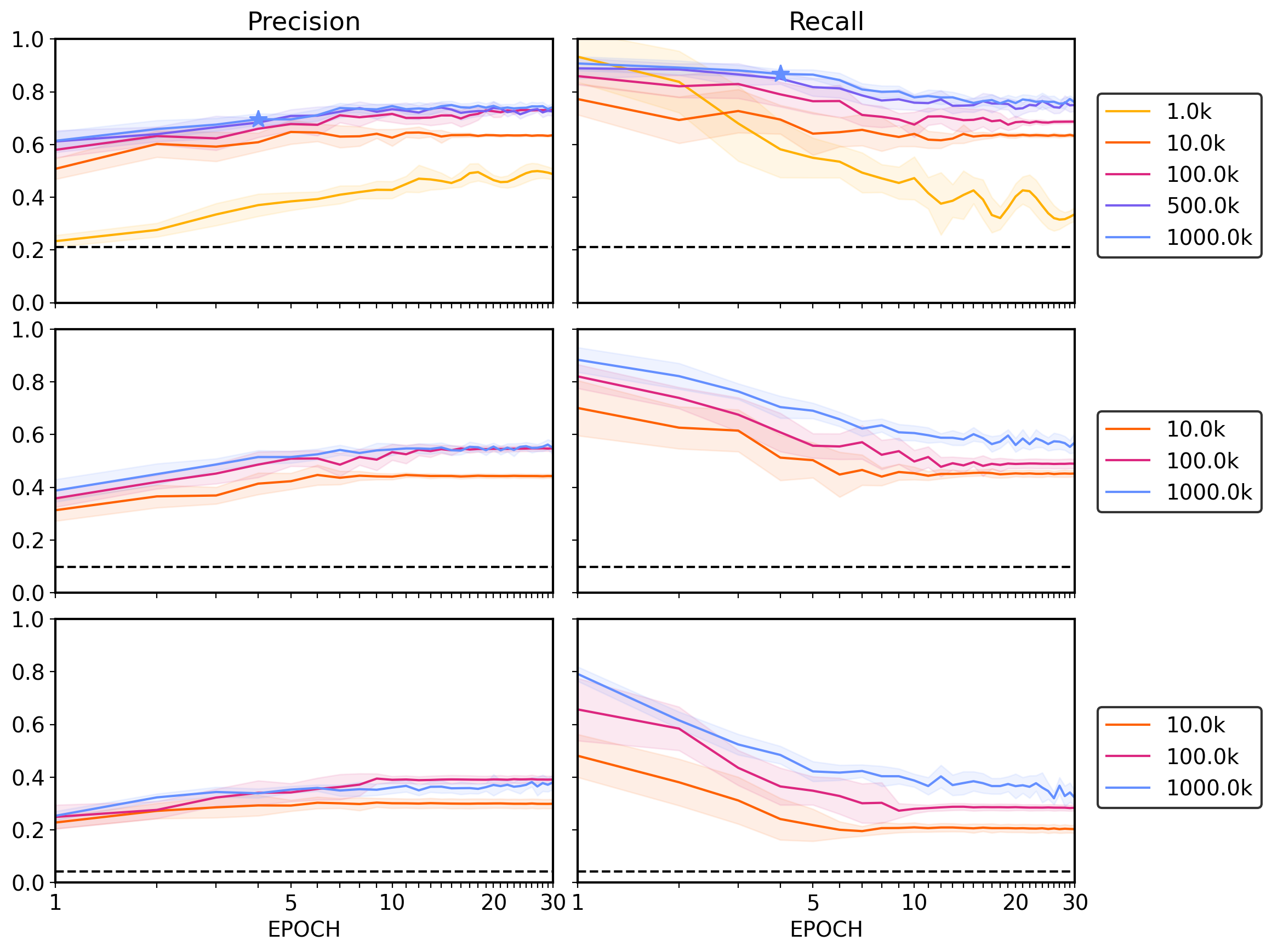}
   \caption{Top: Precision and recall results for models trained on data sets of varying sizes with $D=5$. The dashed black line is the predicted recall and precision from the climatology computed using the largest data set. The blue stars indicate precision=0.70 and recall=0.87. Middle: Same results for $D=7$. Bottom: Same results for $D=9$. Fewer curves are displayed for $D=7$ and $D=9$ for the sake of clarity.}
   \label{fig:Fig5:CNN_complexity}
\end{figure}

The top row of Fig.~\ref{fig:Fig5:CNN_complexity} shows the precision and recall evaluated on the test data  for  varying training data sets for $D=5$. 
Both the precision and recall metrics are plotted starting from the end of Epoch 1 (the leftmost point on the horizontal axis of Fig.~\ref{fig:Fig5:CNN_complexity}); 
From Epoch 2 to Epoch 10, the precision increases, chiefly reflecting a decrease in the false positive rate, as the CNN  becomes better at discriminating between persistent and non-persistent flow configurations. At the same time, the recall slowly decays: the false negative rate rises slightly as the network becomes more conservative and less likely to over-predicting persistent cases.  Except for the low data regime (1k days), the performance of the CNN asymptotes after approximately 10 epochs where the precision and recall are approximately equal, but this is not necessarily the ideal stopping time~\cite{Bouchet2023}.

Applying the definition of best performance in Eq.~\eqref{eq:bestperformance}, the ``best" CNN is obtained by training on the full data set of 1000k days for 4 epochs, indicated by the star in Fig.~\ref{fig:Fig5:CNN_complexity}. It achieves precision of 0.70 and recall of 0.87, exhibiting significant predictive power over the climatological mean prediction (the black dashed line with value 0.21). Therefore, we use it for further analysis in Sec.~\ref{sec:XAI}.

All of our CNNs significantly outperformed the climatological mean prediction for any amount of data or training length.
Interestingly, although the best performance is always realized with the longest trajectory of 1000k days, the sensitivity of precision and recall to the training data size is different. For $D=5$ events, the precision improves with more data up to 100k days (equivalent to approximately 1000 winters), after which additional data does not lead to much improvement. The recall, however, is more data-hungry; its performance continues to improve until data reaching 500k days, equivalent to 5 millennia of winter data.  This reflects the fact that more data continues to help the CNN avoid missing events after its ability to limit false positive forecasts has saturated.

Fig.~\ref{fig:Fig5:CNN_complexity} also shows the results for higher persistence thresholds, $D=7$ and 9.  These thresholds correspond to rarer evnts, and even with the longest trajectory of 1000k days, the precision and recall curves suffer for two reasons.  First, as seen from Tab.~\ref{tab:MMdataset}, the number of positive events drops, effectively limiting the data set almost by a factor of 5 for the most extreme $D=9$ cases.  More importantly, however, it simply becomes harder to discriminate rare events as the data set becomes more imbalanced: less than 1 in 10 nascent blocking states will evolve into a 7 days block, and less than 1 in 20 into a 9 day blocking event.  Without our efforts to overcome this imbalance, a network can classify almost all events correctly by never predicting a persistent case.

Despite the difficulties, the CNNs still show some skill in rare event forecasting.  Given the full 1000k dataset, for $D=9$ the precision and recall converges to about 0.35. While this is only half the values achieved by the CNN in the $D=5$ case, this is almost 10 times the climatological values of precision and recall in that case.  As with the D=5 cases, we found that the recall for $D=7$ and 9 suffers more than the precision when the data set shrinks: with less events to learn from, the CNNs become more conservative and less likely to call an event.  The recall depends on the false negative rate, thus appears more sensitive to class imbalance.  More data gives the network more true positive cases to learn from, appearing to help overcome this challenge.

The low precision and recall values for smaller data sets (1k and 10k) does not bode well for training our CNN on ERA5 data, 
which will be discussed in detail in Section~\ref{sec:transferlearning}.  For $D=5$, there are 273 nascent blocked states in the ERA5 record, 84 of which persist into blocking events (see Table \ref{tab:ERA5dataset}).  This data amount falls between our 1k and 10k cases where data clearly limit performance.  Consistent with our experience with the MM model, achieving a high recall is the most difficult with limited data, and it is with this metric that transfer learning will have the largest impact.

\section{Feature analysis: What is our CNN using to predict blocking events?}
\label{sec:XAI}
Before turning to forecasting in the realistic data regime, we ask what our best CNNs have learned to make these forecasts.  Explainable Artificial Intelligence (XAI) is an array of techniques used to try to gain some understanding of the basis on which neural networks make predictions
~\cite{linardatos2020explainable}. In this section, we use SHapley Additive exPlanation (SHAP) value analysis to dissect the contributions of different atmospheric pressure levels and geographic areas that our CNN is using to make its predictions. We further construct a sparse model using the identified important features as inputs to quantitatively justify their relative importance in the prediction process.
\subsection{Method}
 SHapley Additive exPlanation (SHAP) values, introduced by \citeA{lundberg2017unified} and~\citeA{shrikumar2017learning}, draw inspiration from Shapley values in game theory~\cite{ShapleyGametheory}.  In the domain of weather and climate science, SHAP values have found broad use, with applications ranging from Earth System model error characterization~\cite{silva2022using} to drought forecasting~\cite{DIKSHIT2021100192}.

Intuitively, given a function $f:\mathbb{R}^d\to\mathbb{R}$ (such as the conditional probability function $q$ in Eq.~\ref{eq:q}), SHAP assigns an importance value $\phi_i$ to each feature $x_i$ of the argument $\boldsymbol{x}\in\mathbb{R}^d$, which combine additively:
\begin{align}f(\boldsymbol{x})=\mathbb{E}[f(\boldsymbol{x})]+\sum_{i=1}^d\phi_i(f,\boldsymbol{x}).
\label{shap-def}
\end{align}
With no knowledge of $\boldsymbol{x}$, the optimal prediction of $f$ (in a mean-square sense) is the climatological average over the distribution of $\boldsymbol{x}$: $\mathbb{E}[f(\boldsymbol{x})]$ .  SHAP values quantify how much is gained beyond this baseline by incorporating information from each component $i$ of $\boldsymbol{x}$. The SHAP values $\phi_i (f,\boldsymbol{x})$ are unique for each sample of $\boldsymbol{x}$, but features $i$ for which $\vert \phi_i(f,\boldsymbol{x})\vert$ are large for most $\boldsymbol{x}$ (that is, a large SHAP value on average) can be singled out as important, or useful, for the prediction of $f(\boldsymbol{x})$. SHAP values possess  advantageous theoretical properties as well, and we refer the reader to \citeA{lundberg2017unified} for a detailed theoretical analysis. In this study, SHAP values are computed using the Python package Deep SHAP. The function $f(\boldsymbol{x})$ is taken as the estimated conditional probability $\hat{q}(\boldsymbol{x})$ computed by the CNN, i.e., the probability, according to the CNN, that the blocked state will extend $\geq D$ days, leading to a blocking event.

\subsection{Results}
\begin{figure*}
    \centering
    \includegraphics[width=1.0\textwidth]{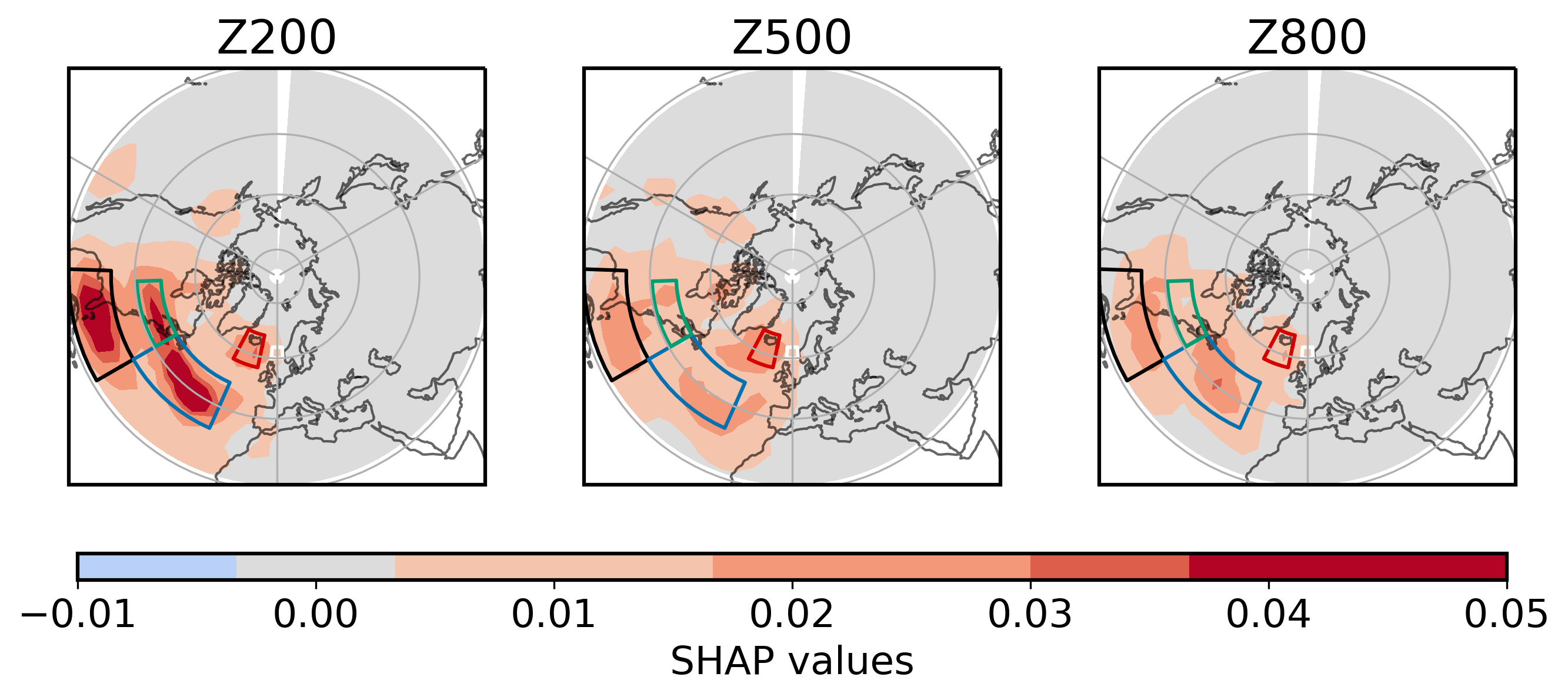}
    \caption{Composite maps of SHAP values, $\overline{\phi}$, of geopotential height at 200, 500, and 800 hPa, for true positive cases, i.e., when the CNN accurately forecasts a persistent blocking event. The unit is the probability of a positive forecast per feature (see equation \ref{shap-def}), indicating the average incremental increase in the CNN's  confidence that the nascent blocked state will evolve into a persistent blocking event, given knowledge of $Z$ at at a given location and pressure.  The boundaries of the most important regions learned by the CNN are marked by solid lines and denoted region 1 (Florida, black), region 2 (north Atlantic, blue), region 3 (northeastern North America, green) and region 4 (Iceland, red) . }
    \label{fig:interpretableAI}
\end{figure*}

Fig.~\ref{fig:interpretableAI} shows the composite of SHAP values for true positive data. Because few nascent blocks persist for $D=5$, 7, or 9, the climalogical probability of a persistent event $\mathbb{E}[\hat{q}(\boldsymbol{x})] = 0.21$, $0.096$, and $0.044$, respectively. For our CCN to call a positive event, we require the conditional forecast probability $\hat{q}(\boldsymbol{x})$ to be larger than 0.5.  Hence a positive (negative) value of $\phi_i (\hat{q},\boldsymbol{x})$ indicates that knowing the geopotential height anomaly at this level and location increases (decreases) the likelihood of a positive event. Therefore, the shading in Fig.~\ref{fig:interpretableAI} can be interpreted as the average influence of each grid point for the CNN to successfully predict a long-lasting blocking event.

The SHAP composite is approximately uniformly non-negative because it is based only on true positive events: additional information should always increase the forecast probability. This indicates that the CNN has been well-trained to only use geopotential height information that improves the blocking event probability, and suggests it has identified robust features that herald a persistent block.  A composite based on true negative cases (not show), reveals similar patterns, but of the opposite  sign.

The first thing to notice is that anomalies upstream from the blocking region (to the west) are more valuable for predicting the persistence of the blocked state. Moreover, the commonality among different pressure levels reflects the relatively barotropic nature of the MM model. In general, however, the CNN prediction relies most on the upper level flow (200 hPa).

The SHAP values emphasize four distinct regions in a quadrupole arrangement to the west of the Atlantic blocking region, as marked in Fig.~\ref{fig:interpretableAI}. We chose these regions to encapsulate high SHAP values using the following algorithm: after objectively identifying regions where SHAP values exceeded a set threshold, we defined boundaries by hand with the goal of enclosing these regions across all three levels within the smallest encompassing rectangle. While part of the goal of choosing these regions was to build a sparse predictor in the next section, they give us physical insight on their own.

\begin{figure*}
    \centering
    \includegraphics[width=1.0\textwidth]{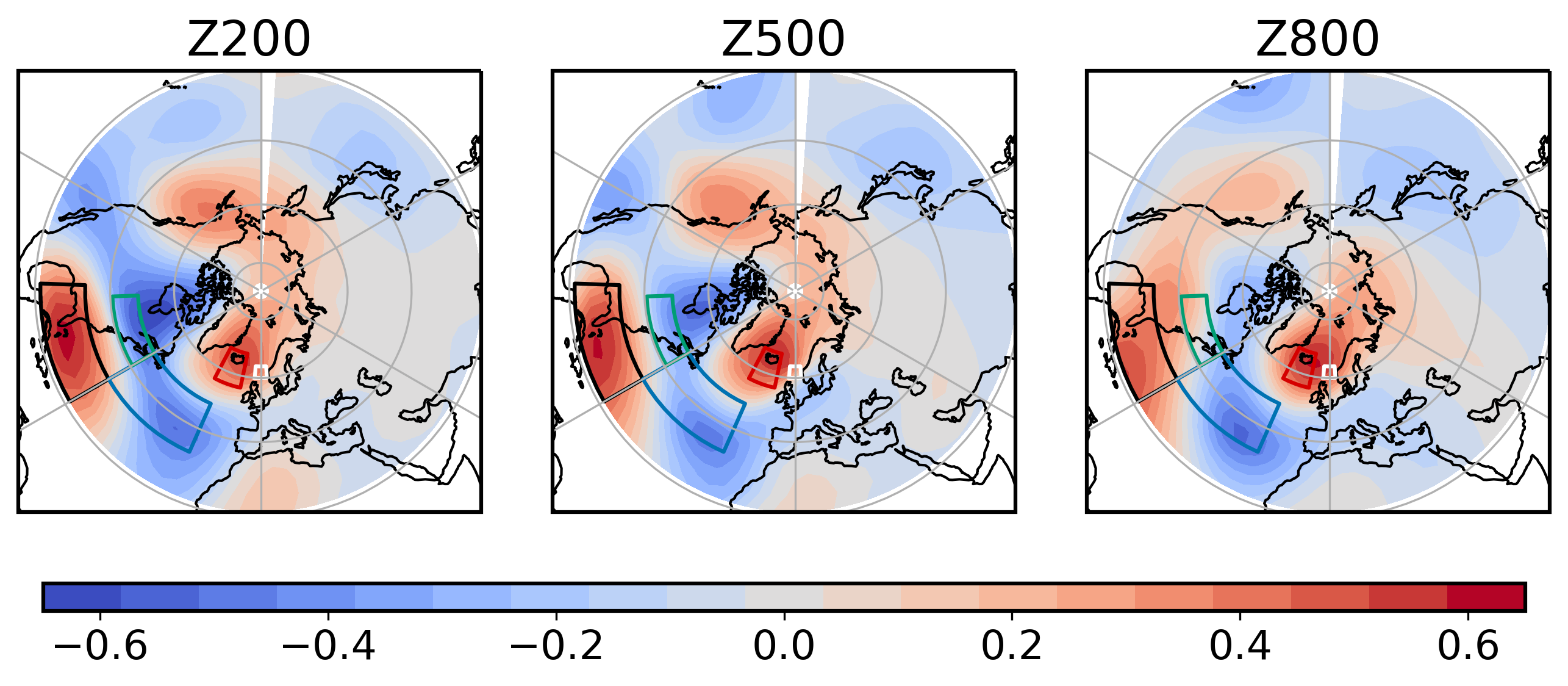}
    \includegraphics[width=1.0\textwidth]{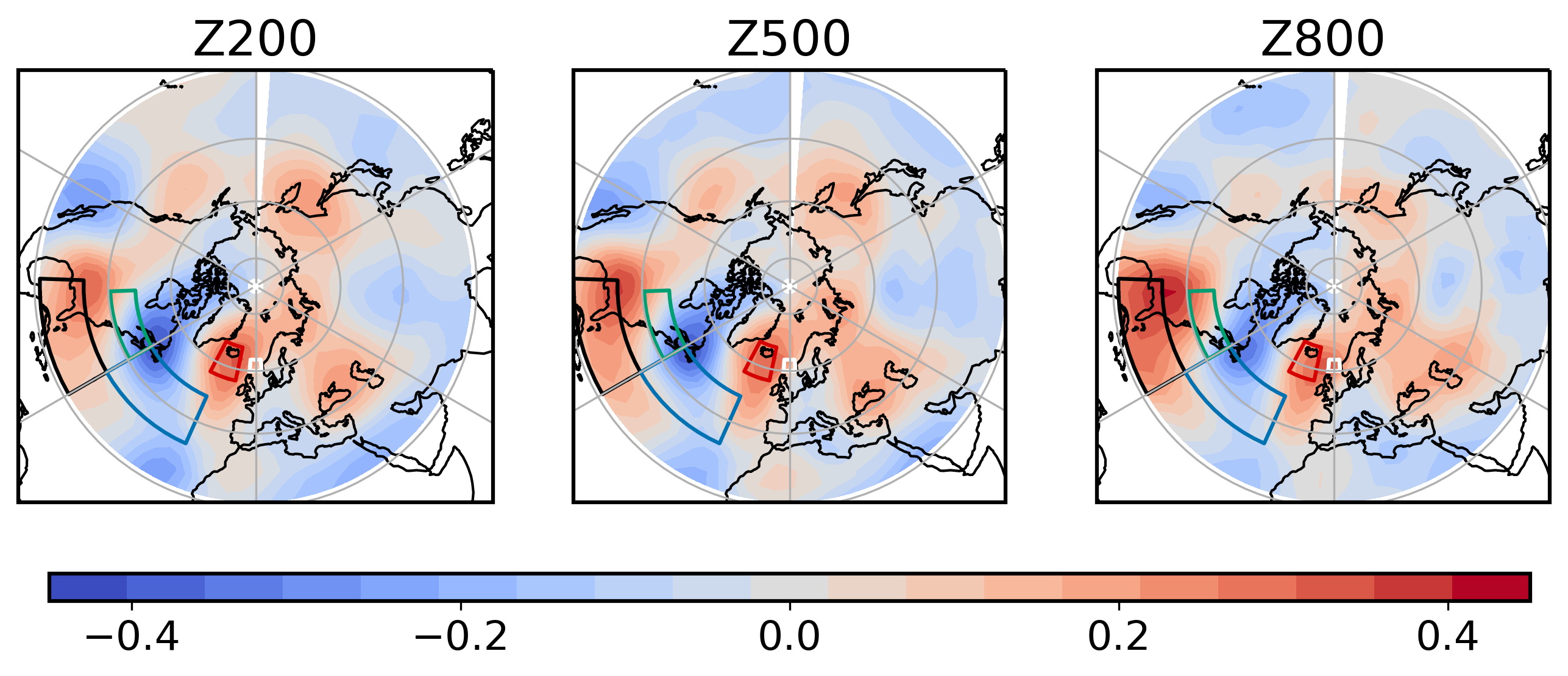}
    \caption{Average states of nascent blocking states that evolve into persistent blocking events ($T=1, y=1$) of (top row) MM dataset and (bottom row) ERA5. The colorbar represents values of   geopotential height anomalies normalized by the standard deviation at each location and height.}
    \label{fig:composite_maps_ERA5_MM}
\end{figure*}

The meaning of the SHAP values can be more easily interpreted with the aid of composites of the true positive events (Fig.~\ref{fig:composite_maps_ERA5_MM}), which show us the sign of anomalies that favor persistence.  Positive geopotential anomalies in region 1 (black, centered over Florida) and 4 (red, over Iceland, just east of the blocking region itself) at the onset of blocking indicate to the CNN that a block will persist, while negative anomalies over Regions 2 (blue, North Atlantic Ocean) and 3 (green, northeast US) also favor persistence.

Regions 2 and 4 project onto opposing centers of action of the North Atlantic Oscillation (NAO).  They indicate that a more negative NAO state at the onset of blocking increases the likelihood of a persistent block.  Previous studies have also found that blocks tend to be more persistent when the NAO is negative \cite{DynamicalFeedbacksBarnes2010}.  While a blocking pattern off Europe projects weakly onto the NAO itself, SHAP analysis indicates that the wider structure of the pattern is important.  Regions 1, 3, and 4, on the other hand,  appear to be part of a wave train arching southwest from the blocking region. Their importance suggests that downstream development of a wave packet propagating along the jet stream helps drive persistent blocking events in the North Atlantic.

\section{Building a sparse model: Logistic regression}
\label{sec:sparsemodel}

\begin{figure}
    \centering
    \includegraphics[width=0.52\columnwidth]{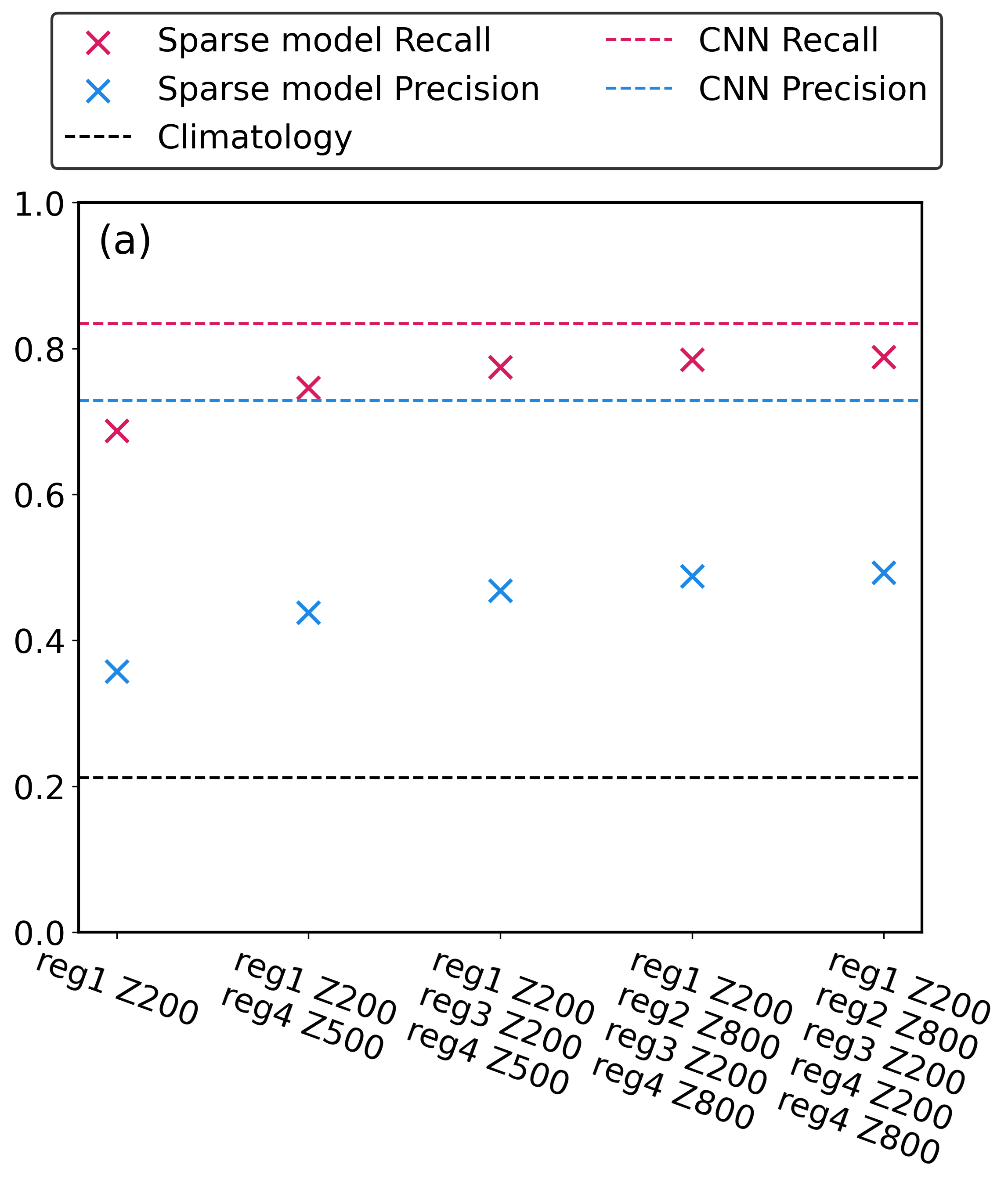}
    \includegraphics[width=0.47\columnwidth]{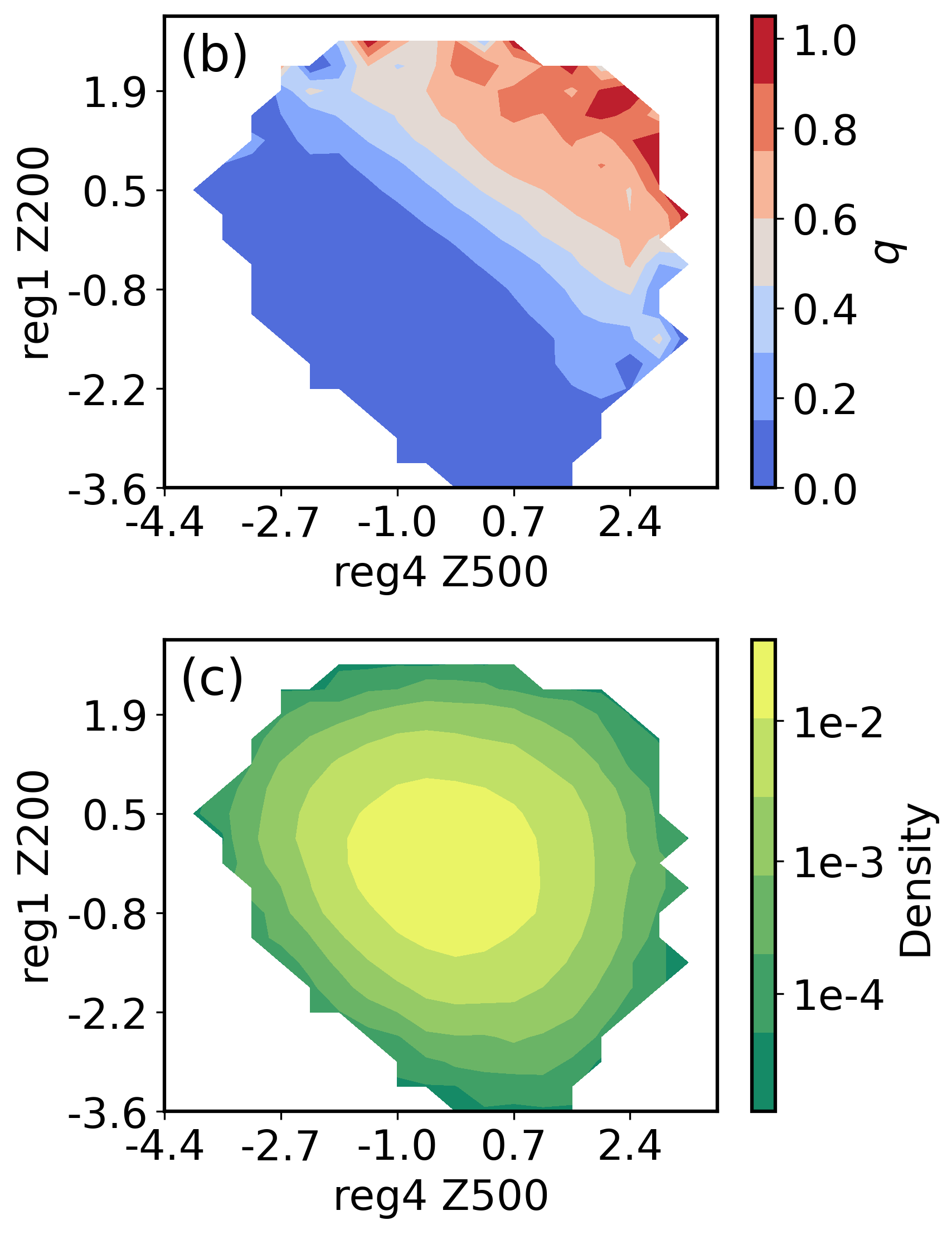}
   \caption{(a): Sparse model predictive skill on the test data set. The horizontal axis represents the dimension $d$ of the sparse model from 1 to 5, with labels showing the combination of variables (``R1'' = ``region 1'') that achieves the best predictive skill among all combinations of $d$ variables. The $(+)$ and $(-)$ indicate the sign of the coefficient before the variable in the logistic regression. (b) Conditional probability of a persistent block, $q$, as a function of mean normalized geopotential height anomaly at 200 mb over region 1 and at 500 mb over region 4 (the second column of (a)). (c) The marginal density (likelihood of observing these anomalies) as a function of the same variables. Densities below $10^{-5}$ are cut off. }
   \label{fig:AI}
\end{figure}

To substantiate the importance of the regions highlighted by the CNN in prediction, we constructed a sparse model. The success of this model reveals that a small set of well-chosen variables and a model with a simple structure can recover a sizeable portion of the predictability. We computed the local mean of $Z200$, $Z500$, $Z800$ for each of the four rectangles shown in Fig.~\ref{fig:interpretableAI}, resulting in 12 time series. We then applied logistic regression with different combinations of these 12 features. The results for the sparse models with the best predictive skill within models of 1 to 5 dimensions on the test set are illustrated in Fig.~\ref{fig:AI}(a). The horizontal axis denotes the variable combinations that achieve the predictive scores shown in the figure.

We draw three key conclusions from Fig.~\ref{fig:AI}(a). First, to predict the persistence of a blocked state, the best one-dimensional feature is Z200 in region 1, upstream over Florida and the Gulf, not Z500 in region 4, the $Z-$field nearest to the blocking region we focus on. Second, the combination of Z200 in region 1, Z500 in region 4 forms a two-dimension model (shown in Fig.~\ref{fig:AI}(b)) that already recovers a recall value of 0.75 -- it captures three quarters
of all blocking events --  with a precision of 0.44, twice the climatological rate. The precision and recall of the full CNN, however, are 0.87 and 0.70. This leads us to the third key message: the large discrepancy in precision between CNN and logistic regression. Even with 5 predictors, the precision of our sparse model is only 0.5.

The poor precision indicates that the sparse model makes too many false positive predictions.  This could suggest that the decay of the Atlantic blocked state is a more nonlinear dynamical phenomenon, which cannot be modeled as a simple linear statistical model. A CNN can capture these nonlinearities more effectively than sparse regression, which is consistent with previous research which found North Atlantic blocks are associated with nonlinear processes~\cite{NonlinearEvans2003}. It could also indicate that more subtle features outside these 4 centers (and variation within these regions) are  important.  Fig.~\ref{fig:interpretableAI} indicates that the CNN uses information across all of the North Atlantic, eastern North America, and even off the west coast of the US, to make skillful predictions.   

To explore the effectiveness of the two-dimensional sparse model,  we visualized the conditional probability of a block persisting, $q$, projected onto this simple subspace (shown in Fig.~\ref{fig:AI}(b)).  Introduced in Eq.~\eqref{eq:q}, $q$  quantifies the probability that the system will evolve into a persistent blocking event before the flow becomes unblocked.  For example, the lightest pink region, corresponding to $q\approx 0.5$ indicates that if, at the onset of blocking, $Z$ at 200 hPa over region 1 (Florida) is particularly high or $Z$ at 500 hPa in region 4 (Iceland) is abnormally high, the system has a roughly 50\% chance of evolving into a persistent block, more than double the climatological rate of 21\%.   In the red region at the top right, where both of these regions exhibit abnormally high pressure, the odds of a persistent block increase to near 100\%.

Fig.~\ref{fig:AI}(c) shows the likelihood of observing these Z200 and 500 anomalies.  Most often, the system exists in the middle of the diagram, where the probability of a blocking event hovers around the climatological value or below.  The most likely state that exhibits a high chance of a block lies along the diagonal from the upper left to the lower right, with moderately high Z200 and 500 anomalies.  The states in the top right corner, for which a persistent block is nearly certain, are very rare.    

The sparse models suggest physical links between blocking events and the upstream flow. The Atlantic blocking region lies at the end of the Atlantic storm track~\cite{michelangeli1998dynamics}.  Persistent blocks, at least in the MM model, are favored when there is enhanced wind off the east coast of the US (high pressure over Florida, region 1) and low pressure over regions 2 and 3 (which are highlighted in the higher dimensional sparse models).  This displaces the climatological winds upstream of the blocking region equatorward.  This will modify the input of storm activity into the blocking region, consistent with prior studies that have highlighted the relation between the storm track and blocking events~\cite{zappa_linking_2014, yang_influence_2021}. 

\section{Extending to ERA5 Using Transfer Learning}
\label{sec:transferlearning}
Given sufficient data, it was possible to construct a CNN that skillfully forecasts the maintenance of blocking events in the MM model. ERA5 December, January and February (DJF) data from 1940-2022 exhibit only 273 nascent blocked states in our Atlantic region of focus.  A significant degradation in performance was evident in Fig.~\ref{fig:Fig5:CNN_complexity} when we restricted the amount of training data from the MM model, the drop unfortunately occuring  in the data regime available in reanalysis.   The curve associated with the trajectory of 10k days (699 nascent blocked states) plateaued at lower values for both the precision and recall.  With only 1k days (63 nascent blocked states) performance was poor, and the learning unstable, oscillating significantly across epochs.

The class imbalance between $Y=0$ and $Y=1$ adds to the difficulty (see Tab.~\ref{tab:ERA5dataset}), particularly when longer blocks are considered. An extreme example is the set of blocking events that last $\geq$9 days: there are only 18 such events ($Y=1$) in the reanalysis record out of 273 data points. Such a small sample of positive data can hardly support any meaningful training, and makes it impractical to get meaningful uncertainty bounds on performance.
In a standard training-test data split with a ratio of 90:10, only around 2 positive events typically fall in the test set, making it challenging to robustly assess the skill.

When training on the limited number of events in the reanalysis, a CNN can more easily suffer from overfitting, where the network uses `noise' (unrelated features) to classify blocking events.  Overfitting can be diagnosed when the performance on the test set diverges from the training set.   Yang and Gerber (submitted) found that the oversampling strategy used so far in this study was more prone to overfitting than a weighted loss function strategy \cite{johnson2019survey}.  With this latter strategy, one emphasizes the rare class (in our case, positive events) by increasing its weight in the loss function. In our remaining experiments, we weighted positive and negative events inverse to their occurance rate. 

\begin{table}
\centering
\begin{tabular}{ c c c } \hline
         Threshold&$Y=1$& $Y=0$\\
         \hline
         $\geq 5$ days&  84& 189\\
         \hline
         $\geq 7$ days&36&237\\ \hline
 $\geq 9$ days& 18&255\\ \hline
\end{tabular}
    \caption{The statistics of ERA5 dataset in 1940-2022 DJF  with $T=1$.}
    \label{tab:ERA5dataset}
\end{table}

\subsection{Direct training}

The scarcity of events makes direct training (DT) on ERA5 blocks challenging. In our study of the MM model data, we had the luxury of a large test data set (which we intentionally kept the same for fair comparison of the different CNNs), even for the case with only 1k training days. For ERA5 data, we use cross validation~\cite{Goodfellow-et-al-2016} to make the best use of the smaller dataset.  The limited number of states were partitioned into training and test sets in ratios of 90:10; we also tried 80:20, and the results were similar (not shown).  These splits were chosen to balance two difficulties: a small training set can prevent robust learning, while a small test data set limits accurate evaluation, even for a well-trained model.

To proceed, we first reduced the resolution of the ERA5 data to a comparable size of the MM output, considering geopotential height on the same three levels at the same coarse resolution.  Reducing the resolution allowed us to use the same CNN architecture, and made transfer learning possible (as discussed below).  It also helped avoid overfitting, reducing the number of input variables relative to the number of events.
Then we created the test-train splits, yielding 10 cross validation sets with distinct test events.  Finally, for each test-train split, we trained and evaluated 10 CNNs, where variations were confined to random weight initialization and shuffling of training data.

Providing meaningful uncertainty on the precision and recall statistics from direct training, shown in the left column of Fig.~\ref{fig:transferlearning_1}, is challenging.  As the 10 CNNs trained on each train-test split are not independent and identically distributed (IID), we first average the skill scores within each split.  The 10 test sets, however, can be viewed as IID samples. The solid lines and shades respectively represent the mean and two-standard deviation bounds of the precision and recall, as a function of epoch, across the 10 splits.

For 5 day blocks, a CNN trained by DT can beat the climatological forecast, albeit only modestly.  Given the small testing data set (27 nascent blocks, of which roughly 8 persist into events), it is important not to put too much stock in the best possible performing network, for CNN can get lucky on a small size of samples. The average performance more reliably quantifies the potential skill.  On average, a CNN can achieve a precision of approximately 0.45: when it calls a persistent blocking event, 4-5 out of 10 times it is correct, as compared to about 3 of 10 in the climatology. The recall was modestly better, the network only missing 4 of 10 actual events, while a climatological forecast would miss 7 of 10.

We also explore 7 day events, where only 13\% of nascent blocks evolve into 7+ day events.  Again, the average CNN modestly beat the climatological forecast in terms of precision: 1/5 of the cases it calls evolve into persistent events, roughly double the success rate by a guess with a Bernouli random variable.  The recall was initially deceptively high (the network captured 5 of 10 blocks), but this skill rapidly decreased with training.  This was due to the fact that CNNs at early stages of DT call too many events. As it trains further, it reduces the forecast rate, declaring fewer false positives at the expense of missing more events.

\subsection{Transfer learning}

Transfer learning (TL) has found broad application in atmospheric science, such as detecting gravity waves~\cite{10062031}, improving extreme heatwave forecast in climate model~\cite{jacques2022deep},  subgrid-scale (SGS) models~\cite{subel2021data}, image restoration ~\cite{photonics9080582} and parameter retrieval from raw dew point temperature profiles~\cite{8518097}.

TL involves pre-training a model on a larger dataset that is similar to the dataset of interest (source domain), then fine-tuning the model on the smaller target dataset (target domain). This approach is particularly beneficial when labeled data for the target task is limited, as it allows the model to exploit learned features and representations from the larger dataset to enhance its performance on the smaller dataset. With this strength, TL has shown its power in forecasting, combining the data from a climate model~\cite{Prediction2020} or a dynamical model \cite{Mu2020}  with the observational record to improve  medium-range weather forecasting and ENSO prediction.

In this section, we applied TL to leverage our MM dataset to predict events in the reanalysis data. As a quasi-geostrophic model, MM has complexity between full climate~\cite{Prediction2020} and low order~\cite{Mu2020} models used in previous transfer learning studies. The overall process is to first `pre-train' a CNN the MM model dataset, learning to capture the characteristic features of blocking. 
While significantly simplified, the MM model is skillful in representing atmospheric variability~\cite{lucarini2020new}, but more importantly provides extensive positive and negative cases to learn from, supporting optimal CNN training, as demonstrated in Sec.~\ref{sec:cnn}. After pre-training, our CNN is then fine-tuned on the ERA5 dataset, where the weights are modified to account for biases in the MM model, and the parameter scales are calibrated.


In most applications of TL, only the weights in the last few layers of a neural network are fine-tuned on the target domain \cite{yosinski2014transferable,hussain2019study,talo2019application}. Following this convention, we only retrain the last layer of the CNN on ERA5 while keeping the other layers frozen.  This allows the CNN to correct biases it inherits from MM, but not to fall back into the poorly constrained limit we reached with direct training.  We also tried retraining other single layers too, but retraining the last layer performed the best. To avoid
overfitting, we set the learning rate to $1/10$ the learning rate of pre-training.

\begin{figure}
    \centering\includegraphics[width=1.0\columnwidth]{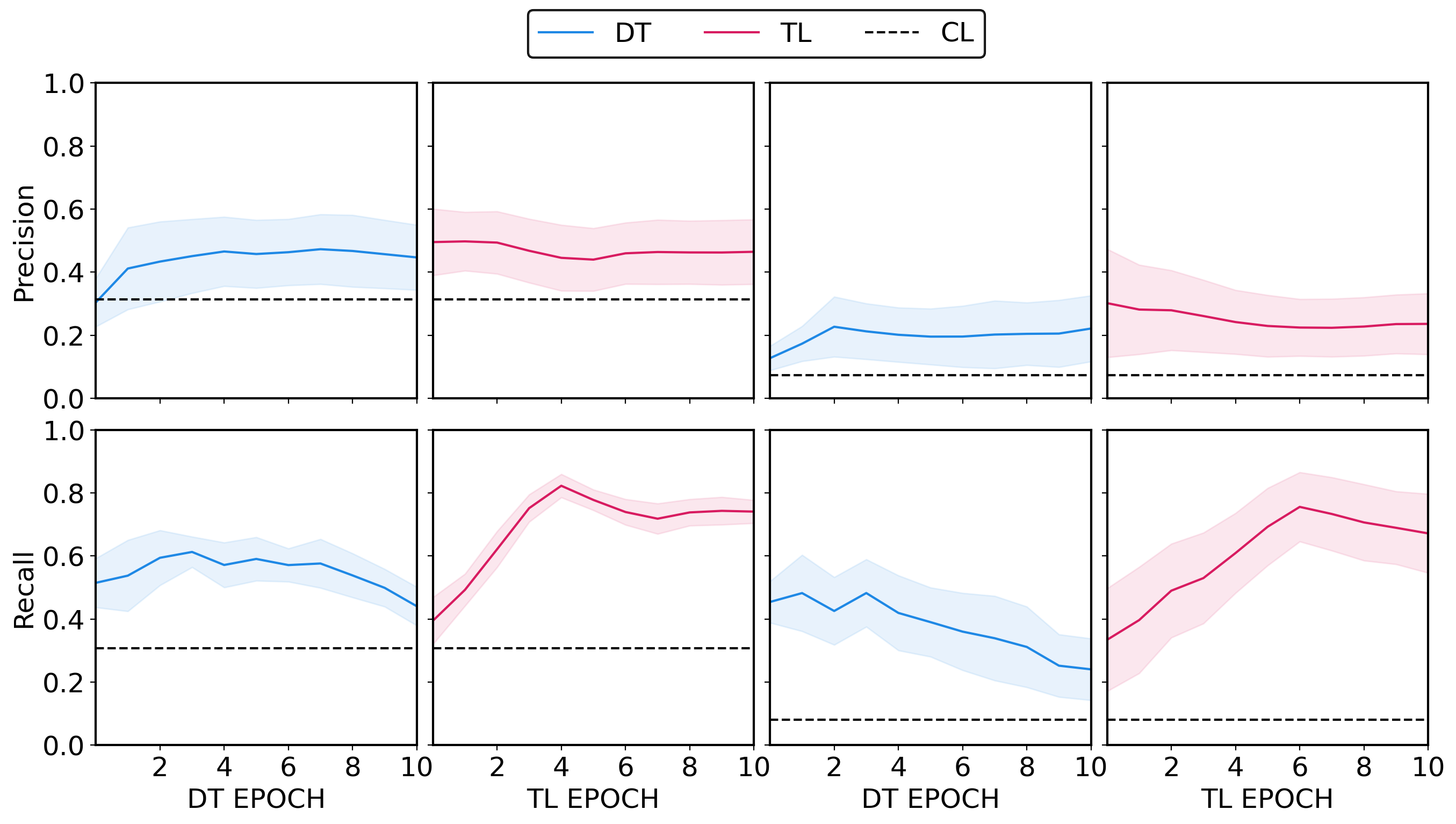}
    \caption{Comparison of CNN forecast skill between direct training (DT, blue) and transfer learning (TL, red). The top row shows the precision and the bottom row, the recall, as a function of training epoch of DT (columns 1 and 3) and fine-tuning epoch of TL (columns 2 and 4). The black dashed line indicates the climatological event rate $p$. The left two columns show the results for $D=5$ (standard blocking events) and the right two columns show the results for $D=7$ (longer blocking events). The shading shows a two-standard deviation uncertainty bound, as detailed in the text.}
    \label{fig:transferlearning_1}
\end{figure}

\begin{figure}    \centering\includegraphics[width=0.8\columnwidth]{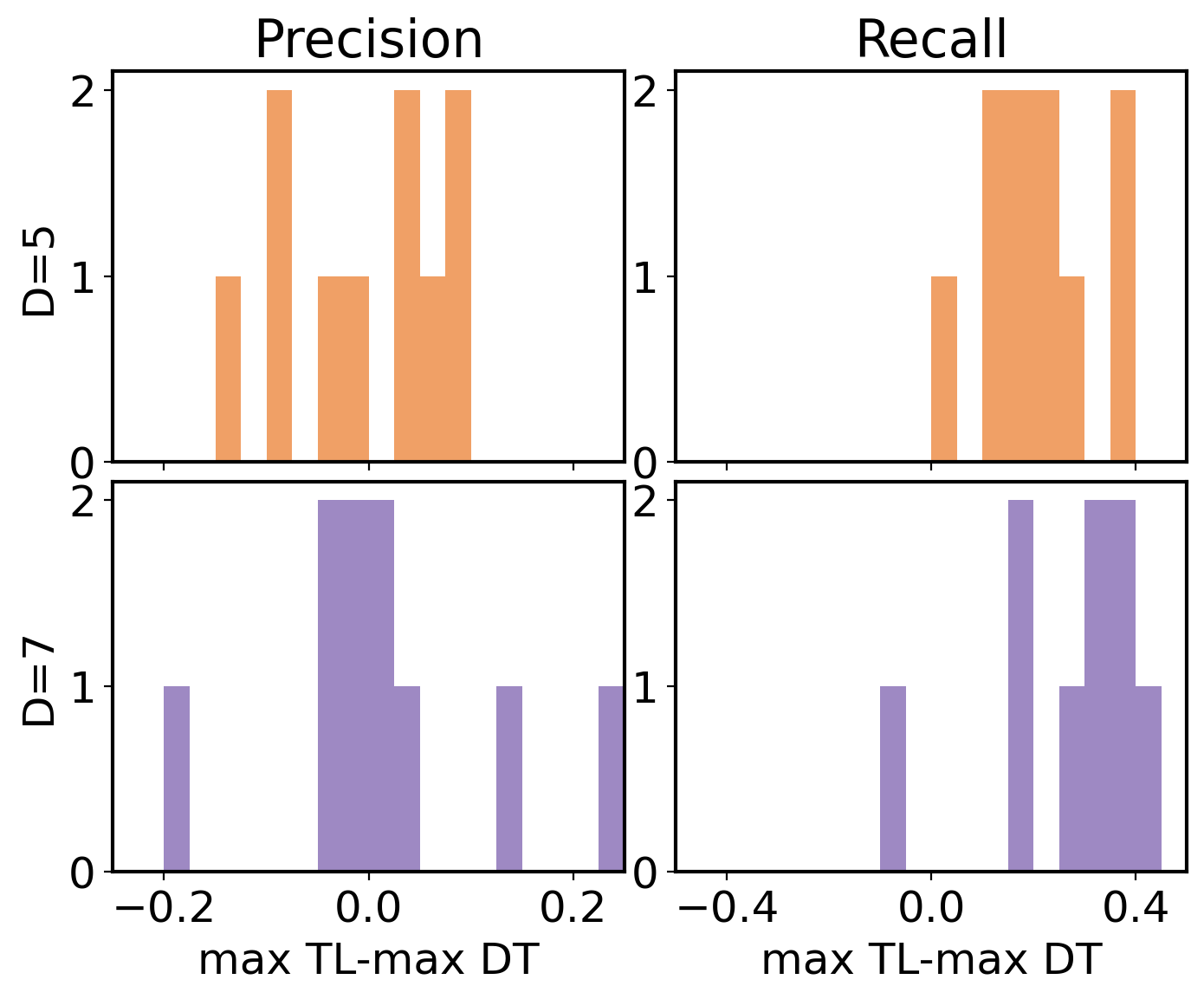}
    \caption{Histograms of the performance gap between the best performing CNNs obtained with transfer learning versus the best performing CNNs obtained with direct training, for (left) precision and (right) recall.  The top panels are for 5 day events and the lower panels are for 7 day events.  ``Best performing'' was determined by stopping the training procedure at the epoch when the best overall balance between high precision and recall was achieved in the mean  (solid lines in Fig.~\ref{fig:transferlearning_1}).  The 90:10 split yields 10 different CNN scores, and the differences between pairs of TL and DT based CNNs, scored on the same test split, are shown.}
    \label{fig:transferlearning_2}
\end{figure}

We tested different lengths of pre-training and then evaluated the performance of the resulting models with the peak precision and recall in the transfer-learning phase. The results show that CNN parameters taken at earlier pre-training epochs show better peak performance after transfer learning (results not shown). This suggests that  overfitting on the source domain cannot be fully corrected by  fine-tuning on the target domain. For the displayed results in Figs.~\ref{fig:transferlearning_1}, \ref{fig:transferlearning_2} and \ref{fig:transfer_learning_shap}, we use a pre-training of 2 epochs for $D=5$, and 1 epoch for $D=7$.  Given the 1000k days of MM integration we had at our disposal, this means that the neural network has explored more than unique 70,000 nascent blocking states (all of them twice, for $D=5$) before seeing any of the 273 events in ERA5.  


We follow a similar procedure as with DT to assess the ensemble-average performance. We pre-train 10 CNNs with the 1000k-day MM dataset; the only differences are due to randomness in the initialization and training data shuffling. We then carry out a 10-fold cross-validation procedure with 90:10 splits: 
for each split, we perform TL fine-tuning on the 10 pre-trained CNNs. We compute the mean precision and recall for each split. The results in the right column of Fig.~\ref{fig:transferlearning_1} show the mean and 2-standard deviation bounds across all the splits. 

Compared to DT, TL begins with a higher precision but lower recall due to pre-training. With additional fine-tuning, the precision stays almost unchanged, while the recall grows markedly.   The network is able to increase the number of events that it can capture (lowering the number of false negatives) with minimal degradation in reliability of its forecast (that is, only slighly increasing the false positive rate).

Uncertainty in the precision is dominated by differences in the true positive events between the splits; consequently, the 2-standard deviation error bounds are comparable for DT and TL.  The recall is less sensitive to differences among the splits, however, and at least for the $D=5$ case, there is noticeably less spread across the splits with transfer learning.  This is understandable because recall, by definition, doesn't depend on the positive rate of the test dataset, which varies a lot for small data sets (around 27 states in each test set after splitting). On the other hand, precision relies on the positive rate of the test dataset, so it has more intrinsic variability.


We still evaluate the overall performance by Eq.~\eqref{eq:bestperformance}. Focusing first on  $D=5$ events, the best mean performance with DT is a precision of 0.45 and recall  of 0.61, which is realized at Epoch 3. With TL, we achieve an average performance with a similar precision of 0.45 and higher recall 0.82 (at Epoch 4). A noticable advantage of TL is the significantly reduced variance in recall compared to DT, indicating TL's superior robustness in prediction, attributed to its enhanced capacity for capturing predictive features. 
For $D=7$ day events, the best mean performance with DT is a precision of 0.21 and recall of 0.48, achieved after 3 epochs. TL, however, achieves a precision of 0.22 and recall of 0.76 at Epoch 6.

To ensure that these gains in recall are statistically significant, we conducted a Wilcoxon signed-rank test~\cite{conover1999practical}.  Fig.~\ref{fig:transferlearning_2} shows histograms of the difference in precision and recall between direct training and transfer learning. For example, each of the 10 values in histogram of $D=5$ is defined for a specific train-test split, evaluated by subtracting the mean precision (recall) of 10 randomly initialized TL models taken at Epoch 4 from the mean precision (recall) of 10 randomly initialized DT models taken at Epoch 3. The spread here stems primarily from the fluctuation in 10 small-size test sets, not uncertainty in the networks due to randomness in training. The values for small-size test sets are taken at the same epoch of the best mean performance.

The average recall with TL surpasses that of DT by 34\% ($p=0.001$) for 5 day events and by over 50\% ($p=0.002$) for 7 day events.  While there is not a significant difference between the TL and DT precision, it is critical that transfer learning was able to improve the recall without sacrificing precision.  One could easily inflate the recall by declaring more positive cases, but without any skill, the precision would suffer and approach the climatological rate.

\subsection{What has transfer learning learned?}
\begin{figure*}
    \centering
     \includegraphics[width=0.75\textwidth]{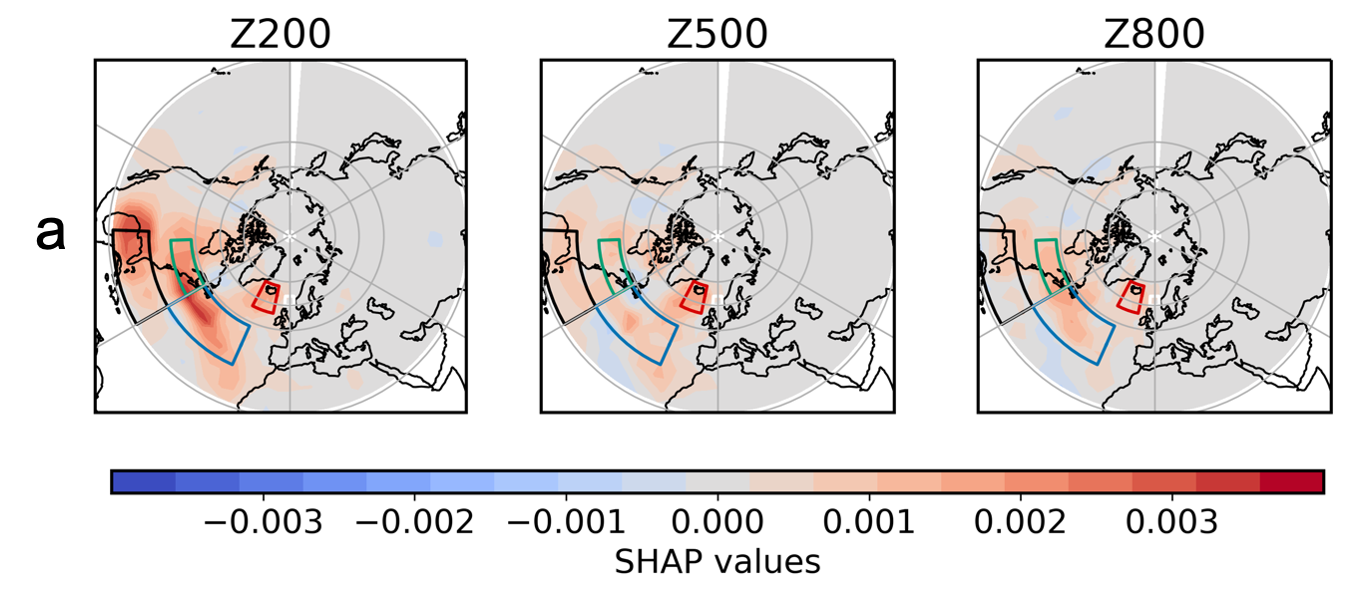}
     \includegraphics[width=0.75\textwidth]{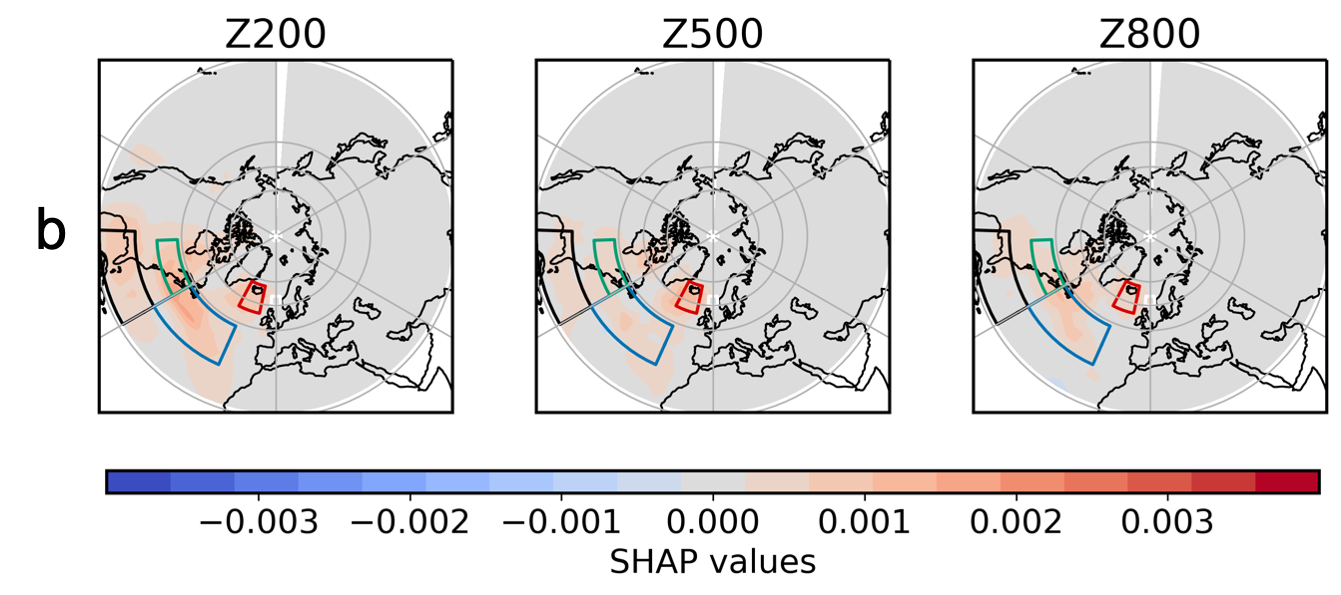}
     \includegraphics[width=0.75\textwidth]{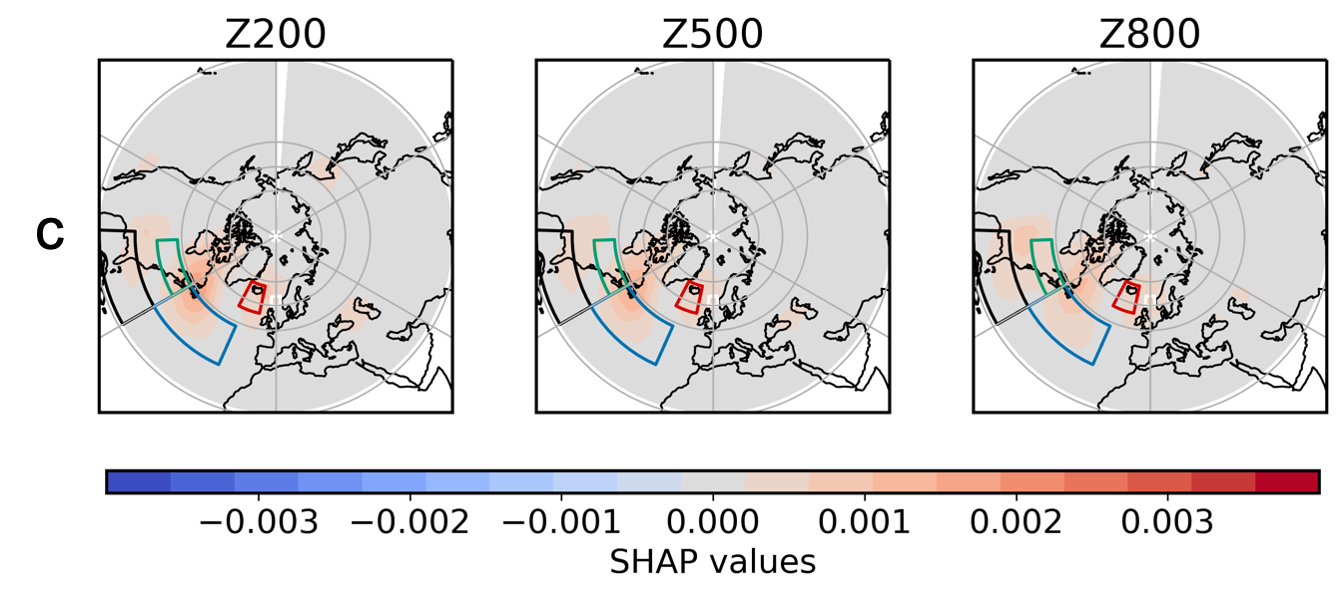}
     \includegraphics[width=0.75\textwidth]{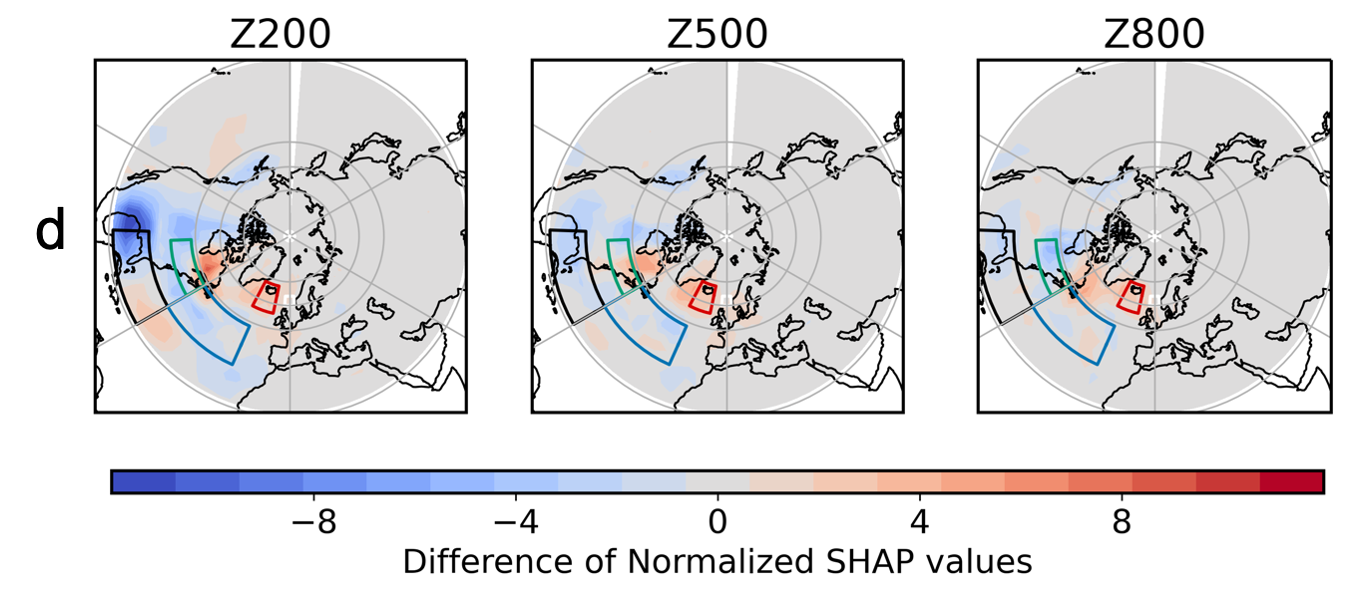}
    \caption{Rows 1 through 4 are composite maps of SHAP values, $\overline{\phi}$, for geopotential height (200, 500, and 800 hPa), averaged over true positive predictions of blocking events in ERA5 by the CNNs listed below.  This is the same quantity shown in Fig.\ref{fig:interpretableAI}, but now applied to ERA5 events. Row $a$ shows $\overline{\phi}^{\text{MM}}$ for the pre-trained CNNs before transfer learning (i.e., networks that have only learned from MM, but applied to ERA5). Row $b$: $\overline{\phi}^{\text{TL}}$ of these pre-trained CNNs after fine-tuning. Row $c$: $\overline{\phi}^{\text{DT}}$ of CNNs directly trained on ERA5 dataset (i.e., networks that never saw the MM events). Row $d$ shows the change in the SHAP values, $\Delta \phi$, between the first two rows, after normalization as detailed in the text.  This quantifies the effect of transfer learning: positive values indicate that information from the region became more important for the prediction, while negative values indicate that anomalies in the region became less important for prediction.}
    \label{fig:transfer_learning_shap}
\end{figure*}

When we show ERA5 events to CNNs first trained on the MM dataset, what exactly is the CNN learning to improve the recall? For example, do the key geographical regions and levels (Fig.~\ref{fig:interpretableAI}) retain the same level of significance? It is reasonable to expect that this might not be the case. In the MM dataset, the duration of the Atlantic blockings could be related to upstream flow, specifically to the structure of the wave train at the blocking onset. The mechanism for blocking in the real world is more complicated, and the correlated pattern may shift, intensify, and/or weaken.  To address these questions, we compare the SHAP values of the pre-trained CNNs when directly applied to ERA5 (i.e., without fine-tuning step) to the SHAP values of the CNN after 4 epochs of fine-tuning, as shown in row $a$ and row $b$ of Fig.~\ref{fig:transfer_learning_shap}.
The most evident difference after fine-tuning is a decrease in the amplitude of the SHAP values. This is because the climatological rate of positive blocking events in ERA5 is higher: almost 1/3 of nascent blocked states persist for 5 days in ERA5, compared to about 1/5 in MM.  As the expected fraction of events is larger, $\hat{q}(\mathbf{x})-\mathbb{E}[\hat{q}(\mathbf{x})]$ from equation (\ref{shap-def}) will be smaller, and the SHAP value increments $\phi_i(\hat{q},\mathbf{x})$ will tend to be smaller.  It is the sum of the SHAP values that build up the probability for a $Y=1$ prediction; for a more likely event, one does not need to build up the probability as much, so fine-tuning quickly adjusts the weights.

To assess the more subtle change in the relative contribution of each feature on the predicted result after transfer learning, we show the difference in the normalized composite map $\Delta \phi$ in row $d$ of Fig.~\ref{fig:transfer_learning_shap}.   
$\Delta \phi$ is defined for each input $i$ (i.e., geopotential height $Z$ at a particular latitude, longitude, and pressure level) by $\Delta \phi_i\equiv \max\left(\frac{\overline{\phi}^{\text{TL}}_i}{\frac{1}{d}\sum_{j=1}^d\overline{\phi}^{\text{TL}}_j},0\right)-\max\left(\frac{\overline{\phi} _i}{\frac{1}{d}\sum_{j=1}^d\overline{\phi} _j},0\right)$. The maximum function is used to avoid spurious negative SHAP values, which should not arise in a composite of true positive events, as discussed in the context of Fig.~\ref{fig:interpretableAI}. The normalization makes the total integral of the SHAP values the same for both cases, so that one can focus on where the CNN is using information, as opposed to the overall reduction of the SHAP values driven by the difference in rates.

The ``normalized'' SHAP values increase mainly in region 4 (the region right around the block), and additionally over Quebec and Atlantic Canada, a region less used for predictions with the MM model. The SHAP values decrease in a relative sense over regions 1 (Florida and the Gulf), 2 (North Atlantic Ocean), 3 (northeastern North America), and central North America. This change in  relative importance reveals a general de-emphasis of the regions farther upstream and an increased emphasis on regions more immediately upstream.  This indicates that while it is still upstream information that is most important for predicting a persistent blocking state in ERA5, the structure and westward extension of the wave train has changed.

For further insight, we compare the SHAP value patterns with a more traditional metric for understanding predictability: composite analysis.  Fig.~\ref{fig:composite_maps_ERA5_MM} shows composite maps of nascent blocks that evolve into persistent events in the MM model and ERA5.  Persistent blocks are associated with wave activity south and west of the blocking region in both the model and reanalysis, but the pattern shifts.  The wave train in MM initially arcs westward before turning southward, with a strong center of high pressure east of Florida, while the wave train in ERA5 arcs more to southwest at first, then further westward.

The SHAP values change over Quebec, capturing this shift in the wave train, but overall the CNN seems to shift to more local information with transfer learning. We speculated that the dry, quasi-geostrophic MM model overemphasizes long range teleconnections. It only captures deformation scale dynamics, and this only at low resolution, and so lacks smaller, local modes of instability, e.g., instability associated with latent heat release due to precipitation, present in our atmosphere. The CNN makes more use of these local features when predicting the persistence of blocks, but still focuses on the upstream flow, consistent with our intuition.

Finally, we contrast the feature importance analysis of the CNN with transfer learning (Fig.~\ref{fig:transfer_learning_shap} row $b$) to that of the CNNs trained only directly on the ERA5 output (Fig.~\ref{fig:transfer_learning_shap} row $c$).  DT struggles to develop nuanced features with limited data. The SHAP values with DT are also more barotropic than those with TL. Moreover, in general, the SHAP values with TL capture finer details across a wider spatial range, while the SHAP values with DT are more localized. Geopotential height anomalies over Iceland, especially in the Z500 map, are more emphasized for TL than DT. The same applies to upstream anomalies over Florida and the Gulf of Mexico in the Z200 map. Additionally, the importance of geopotential height anomalies over the Atlantic, immediately upstream of the target region west of north Africa, is neglected in DT, though it appears in TL. This is closely correlated to the blocking event prediction from the ERA5 composite in Fig.~\ref{fig:composite_maps_ERA5_MM}, which does not show as strong composite Atlantic anomaly as in the MM model.

In summary, the superiority of CNNs trained with transfer learning, as compared to direct training, appears to lie in their ability to leverage learned features from the pre-trained dataset, helping the network to take advantage of information further upstream of the blocking region. In either case the precision is modest: when the networks call an event, the rate of success is at best 50\% higher than a na\"{i}ve climatological forecast.  Pre-training the network, however, has a significant impact on the recall, increasing the forecast rate to capture more events without decreasing the precision.


\section{Conclusion}

The impact of data-driven science on weather and climate science has grown substantially in recent years. In this paper, we suggest two data-driven approaches to help predict and understand atmospheric blocking events. First, given sufficient data, convolutional neural networks (CNNs) are capable of identifying subtle features that differentiate  short-lived blocked states from those that persist for an extended period.  Moreover, XAI methods can provide  insight into what features matter most to this differentiation.  Second, transfer learning has the potential to make data-driven forecasts possible for our atmosphere, making the most of the limited extreme events in the observational record by leveraging insight from longer, albeit imperfect, numerical simulations.

We began in a data-rich regime with the idealized Marshall-Molteni model, showing that a CNN can accurately predict the persistence of North Atlantic blocks in terms of both precision and recall. Leveraging XAI (SHAP feature importance analysis), we identified crucial regions for the prediction of persistent blocked states, given a nascent high-pressure anomaly. Our results suggest that incorporation of both local and non-local features is important for prediction skill.

To validate our discovery, we constructed a two-dimensional model that used only upstream anomalies over Florida and the Gulf of Mexico, and anomalies immediately upstream of the blocking region. The sparse model exhibited precision significantly above the climatological rate and recall nearly as good as the full CNN. It struggled, however, with false positives (and hence exhibited low precision relative to the CNN) which could not be improved  within the log linear logistic regression framework. This suggests the CNN learns non-trivial relations in the upstream flow, extending all the way to the Pacific, to better discriminate between short-lived  and long-lived blocks.

The challenge of conducting direct training on ERA5 data stems from the paucity of available events. Small training and test datasets make training and evaluation difficult.  With the MM model, we observed a systematic degradation in forecast skill when the training data was limited, particularly for the recall statistic.  Through transfer learning, we leverage the abundance of data generated by simplified dynamical models to enhance real-world forecasting. By pre-training a CNN on the MM model dataset and retraining the deepest layer on the ERA5 dataset, the recall was improved by 34\% compared to a CNN developed with direct training alone for 5 day events, and over 50\% for more extreme 7 day events, without any loss of precision.

In addition to advancing predictive skill, transfer learning in combination with SHAP analysis allowed us to compare the predictive features between weather systems in ERA5 and the idealized quasigeostrophic model.  The bottom row of Fig.~\ref{fig:composite_maps_ERA5_MM} reveals biases in the MM model, which appears overly dependent on upstream features over Florida and the Gulf of Mexico relative to blocks in ERA5.   This approach provides a new angle of how a machine learning approach could guide the diagnosis and quantification of model biases.  
This said, the success of transfer learning results underscore the MM model's ability, despite its simplicity, to capture features that are important for predicting the persistence of blocked states in the real world.  We believe that greater strides could be made by pre-training on a more advanced climate model, or even hindcasts in the subseasonal-to-seasonal (S2S) data set \cite{Vitart2017subseasonal,Finkel2023}.

The methods presented here are not limited to the context of blocking events, and can be generalized to the study of other  challenging natural phenomena, especially in scenarios where data may be limited, and the potential influencing factors are complex.  
An immediate future goal is to push further on the physical and dynamical mechanisms that causes the differences in prediction mechanisms for ERA5 and MM model.
Another goal is to adapt the present approach to
investigate the statistical behavior and mechanisms for the onset of the blocking events.


\section*{Open Research Section}
The code for computing SHAP values, transfer learning and producing plots is publicly available in the Github repository at \url{https://github.com/hzhang-math/Blocking_SHAP_TL}.

\acknowledgments
We thank Valerio Lucarini and Andrey Gritsun for sharing their Marshall-Molteni Fortran code.  This work was supported by the Army Research Office, grant number W911NF-22-2-0124. EPG acknowledges support from the National Science Foundation through award OAC-2004572. J. F. is supported through the MIT Climate Grand Challenge on Weather and Climate Extremes, and the Virtual Earth Systems Research Institute (VESRI) at Schmidt Sciences.

\bibliography{references_cap}

\begin{thebibliography}{}

\bibitem [\protect \citeauthoryear {%
Barnes%
\ \BBA {} Hartmann%
}{%
Barnes%
\ \BBA {} Hartmann%
}{%
{\protect \APACyear {2010}}%
}]{%
DynamicalFeedbacksBarnes2010}
\APACinsertmetastar {%
DynamicalFeedbacksBarnes2010}%
\begin{APACrefauthors}%
Barnes, E\BPBI A.%
\BCBT {}\ \BBA {} Hartmann, D\BPBI L.%
\end{APACrefauthors}%
\unskip\
\newblock
\APACrefYearMonthDay{2010}{}{}.
\newblock
{\BBOQ}\APACrefatitle {Dynamical Feedbacks and the Persistence of the NAO}
  {Dynamical feedbacks and the persistence of the nao}.{\BBCQ}
\newblock
\APACjournalVolNumPages{Journal of the Atmospheric Sciences}{67}{3}{851 - 865}.
\newblock
\begin{APACrefURL}
  \url{https://journals.ametsoc.org/view/journals/atsc/67/3/2009jas3193.1.xml}
  \end{APACrefURL}
\newblock
\begin{APACrefDOI} \doi{10.1175/2009JAS3193.1} \end{APACrefDOI}
\PrintBackRefs{\CurrentBib}

\bibitem [\protect \citeauthoryear {%
Berckmans%
, Woollings%
, Demory%
, Vidale%
\BCBL {}\ \BBA {} Roberts%
}{%
Berckmans%
\ \protect \BOthers {.}}{%
{\protect \APACyear {2013}}%
}]{%
berckmans2013atmospheric}
\APACinsertmetastar {%
berckmans2013atmospheric}%
\begin{APACrefauthors}%
Berckmans, J.%
, Woollings, T.%
, Demory, M\BHBI E.%
, Vidale, P\BHBI L.%
\BCBL {}\ \BBA {} Roberts, M.%
\end{APACrefauthors}%
\unskip\
\newblock
\APACrefYearMonthDay{2013}{}{}.
\newblock
{\BBOQ}\APACrefatitle {{A}tmospheric blocking in a high resolution climate
  model: influences of mean state, orography and eddy forcing} {{A}tmospheric
  blocking in a high resolution climate model: influences of mean state,
  orography and eddy forcing}.{\BBCQ}
\newblock
\APACjournalVolNumPages{Atmospheric Science Letters}{14}{1}{34--40}.
\PrintBackRefs{\CurrentBib}

\bibitem [\protect \citeauthoryear {%
Chan%
, Hassanzadeh%
\BCBL {}\ \BBA {} Kuang%
}{%
Chan%
\ \protect \BOthers {.}}{%
{\protect \APACyear {2019}}%
}]{%
Pedramblockingindices}
\APACinsertmetastar {%
Pedramblockingindices}%
\begin{APACrefauthors}%
Chan, P\BHBI W.%
, Hassanzadeh, P.%
\BCBL {}\ \BBA {} Kuang, Z.%
\end{APACrefauthors}%
\unskip\
\newblock
\APACrefYearMonthDay{2019}{}{}.
\newblock
{\BBOQ}\APACrefatitle {{E}valuating {I}ndices of {B}locking {A}nticyclones in
  {T}erms of {T}heir {L}inear {R}elations {W}ith {S}urface {H}ot {E}xtremes}
  {{E}valuating {I}ndices of {B}locking {A}nticyclones in {T}erms of {T}heir
  {L}inear {R}elations {W}ith {S}urface {H}ot {E}xtremes}.{\BBCQ}
\newblock
\APACjournalVolNumPages{Geophysical Research Letters}{46}{9}{4904-4912}.
\newblock
\begin{APACrefURL}
  \url{https://agupubs.onlinelibrary.wiley.com/doi/abs/10.1029/2019GL083307}
  \end{APACrefURL}
\newblock
\begin{APACrefDOI} \doi{https://doi.org/10.1029/2019GL083307} \end{APACrefDOI}
\PrintBackRefs{\CurrentBib}

\bibitem [\protect \citeauthoryear {%
Charney%
\ \BBA {} DeVore%
}{%
Charney%
\ \BBA {} DeVore%
}{%
{\protect \APACyear {1979}}%
}]{%
charney1979multiple}
\APACinsertmetastar {%
charney1979multiple}%
\begin{APACrefauthors}%
Charney, J\BPBI G.%
\BCBT {}\ \BBA {} DeVore, J\BPBI G.%
\end{APACrefauthors}%
\unskip\
\newblock
\APACrefYearMonthDay{1979}{}{}.
\newblock
{\BBOQ}\APACrefatitle {{M}ultiple flow equilibria in the atmosphere and
  blocking} {{M}ultiple flow equilibria in the atmosphere and blocking}.{\BBCQ}
\newblock
\APACjournalVolNumPages{Journal of Atmospheric Sciences}{36}{7}{1205--1216}.
\PrintBackRefs{\CurrentBib}

\bibitem [\protect \citeauthoryear {%
Conover%
}{%
Conover%
}{%
{\protect \APACyear {1999}}%
}]{%
conover1999practical}
\APACinsertmetastar {%
conover1999practical}%
\begin{APACrefauthors}%
Conover, W\BPBI J.%
\end{APACrefauthors}%
\unskip\
\newblock
\APACrefYear{1999}.
\newblock
\APACrefbtitle {{P}ractical nonparametric statistics} {{P}ractical
  nonparametric statistics}\ (\BVOL~350).
\newblock
\APACaddressPublisher{}{john wiley \& sons}.
\PrintBackRefs{\CurrentBib}

\bibitem [\protect \citeauthoryear {%
Davini%
\ \BBA {} D'Andrea%
}{%
Davini%
\ \BBA {} D'Andrea%
}{%
{\protect \APACyear {2020}}%
}]{%
ClimateDavini2020}
\APACinsertmetastar {%
ClimateDavini2020}%
\begin{APACrefauthors}%
Davini, P.%
\BCBT {}\ \BBA {} D'Andrea, F.%
\end{APACrefauthors}%
\unskip\
\newblock
\APACrefYearMonthDay{2020}{}{}.
\newblock
{\BBOQ}\APACrefatitle {{F}rom {CMIP}3 to {CMIP}6: {N}orthern {H}emisphere
  {A}tmospheric {B}locking {S}imulation in {P}resent and {F}uture {C}limate}
  {{F}rom {CMIP}3 to {CMIP}6: {N}orthern {H}emisphere {A}tmospheric {B}locking
  {S}imulation in {P}resent and {F}uture {C}limate}.{\BBCQ}
\newblock
\APACjournalVolNumPages{Journal of Climate}{33}{23}{10021 - 10038}.
\newblock
\begin{APACrefURL}
  \url{https://journals.ametsoc.org/view/journals/clim/33/23/jcliD190862.xml}
  \end{APACrefURL}
\newblock
\begin{APACrefDOI} \doi{https://doi.org/10.1175/JCLI-D-19-0862.1}
  \end{APACrefDOI}
\PrintBackRefs{\CurrentBib}

\bibitem [\protect \citeauthoryear {%
Davini%
\ \BBA {} D’Andrea%
}{%
Davini%
\ \BBA {} D’Andrea%
}{%
{\protect \APACyear {2016}}%
}]{%
davini2016northern}
\APACinsertmetastar {%
davini2016northern}%
\begin{APACrefauthors}%
Davini, P.%
\BCBT {}\ \BBA {} D’Andrea, F.%
\end{APACrefauthors}%
\unskip\
\newblock
\APACrefYearMonthDay{2016}{}{}.
\newblock
{\BBOQ}\APACrefatitle {{N}orthern {H}emisphere atmospheric blocking
  representation in global climate models: twenty years of improvements?}
  {{N}orthern {H}emisphere atmospheric blocking representation in global
  climate models: twenty years of improvements?}{\BBCQ}
\newblock
\APACjournalVolNumPages{Journal of Climate}{29}{24}{8823--8840}.
\PrintBackRefs{\CurrentBib}

\bibitem [\protect \citeauthoryear {%
Dikshit%
\ \BBA {} Pradhan%
}{%
Dikshit%
\ \BBA {} Pradhan%
}{%
{\protect \APACyear {2021}}%
}]{%
DIKSHIT2021100192}
\APACinsertmetastar {%
DIKSHIT2021100192}%
\begin{APACrefauthors}%
Dikshit, A.%
\BCBT {}\ \BBA {} Pradhan, B.%
\end{APACrefauthors}%
\unskip\
\newblock
\APACrefYearMonthDay{2021}{}{}.
\newblock
{\BBOQ}\APACrefatitle {{E}xplainable {AI} in drought forecasting}
  {{E}xplainable {AI} in drought forecasting}.{\BBCQ}
\newblock
\APACjournalVolNumPages{Machine Learning with Applications}{6}{}{100192}.
\newblock
\begin{APACrefURL}
  \url{https://www.sciencedirect.com/science/article/pii/S2666827021000967}
  \end{APACrefURL}
\newblock
\begin{APACrefDOI} \doi{https://doi.org/10.1016/j.mlwa.2021.100192}
  \end{APACrefDOI}
\PrintBackRefs{\CurrentBib}

\bibitem [\protect \citeauthoryear {%
Dole%
\ \BBA {} Gordon%
}{%
Dole%
\ \BBA {} Gordon%
}{%
{\protect \APACyear {1983}}%
}]{%
dole1983persistent}
\APACinsertmetastar {%
dole1983persistent}%
\begin{APACrefauthors}%
Dole, R\BPBI M.%
\BCBT {}\ \BBA {} Gordon, N\BPBI D.%
\end{APACrefauthors}%
\unskip\
\newblock
\APACrefYearMonthDay{1983}{}{}.
\newblock
{\BBOQ}\APACrefatitle {{P}ersistent anomalies of the extratropical {N}orthern
  {H}emisphere wintertime circulation: {G}eographical distribution and regional
  persistence characteristics} {{P}ersistent anomalies of the extratropical
  {N}orthern {H}emisphere wintertime circulation: {G}eographical distribution
  and regional persistence characteristics}.{\BBCQ}
\newblock
\APACjournalVolNumPages{Monthly Weather Review}{111}{8}{1567--1586}.
\PrintBackRefs{\CurrentBib}

\bibitem [\protect \citeauthoryear {%
d’Andrea%
\ \protect \BOthers {.}}{%
d’Andrea%
\ \protect \BOthers {.}}{%
{\protect \APACyear {1998}}%
}]{%
d1998northern}
\APACinsertmetastar {%
d1998northern}%
\begin{APACrefauthors}%
d’Andrea, F.%
, Tibaldi, S.%
, Blackburn, M.%
, Boer, G.%
, D{\'e}qu{\'e}, M.%
, Dix, M.%
\BDBL {}others%
\end{APACrefauthors}%
\unskip\
\newblock
\APACrefYearMonthDay{1998}{}{}.
\newblock
{\BBOQ}\APACrefatitle {{N}orthern {H}emisphere atmospheric blocking as
  simulated by 15 atmospheric general circulation models in the period
  1979--1988} {{N}orthern {H}emisphere atmospheric blocking as simulated by 15
  atmospheric general circulation models in the period 1979--1988}.{\BBCQ}
\newblock
\APACjournalVolNumPages{Climate Dynamics}{14}{}{385--407}.
\PrintBackRefs{\CurrentBib}

\bibitem [\protect \citeauthoryear {%
Evans%
\ \BBA {} Black%
}{%
Evans%
\ \BBA {} Black%
}{%
{\protect \APACyear {2003}}%
}]{%
NonlinearEvans2003}
\APACinsertmetastar {%
NonlinearEvans2003}%
\begin{APACrefauthors}%
Evans, K\BPBI J.%
\BCBT {}\ \BBA {} Black, R\BPBI X.%
\end{APACrefauthors}%
\unskip\
\newblock
\APACrefYearMonthDay{2003}{}{}.
\newblock
{\BBOQ}\APACrefatitle {Piecewise Tendency Diagnosis of Weather Regime
  Transitions} {Piecewise tendency diagnosis of weather regime
  transitions}.{\BBCQ}
\newblock
\APACjournalVolNumPages{Journal of the Atmospheric Sciences}{60}{16}{1941 -
  1959}.
\newblock
\begin{APACrefURL}
  \url{https://journals.ametsoc.org/view/journals/atsc/60/16/1520-0469_2003_060_1941_ptdowr_2.0.co_2.xml}
  \end{APACrefURL}
\newblock
\begin{APACrefDOI} \doi{10.1175/1520-0469(2003)060<1941:PTDOWR>2.0.CO;2}
  \end{APACrefDOI}
\PrintBackRefs{\CurrentBib}

\bibitem [\protect \citeauthoryear {%
Ferranti%
, Corti%
\BCBL {}\ \BBA {} Janousek%
}{%
Ferranti%
\ \protect \BOthers {.}}{%
{\protect \APACyear {2015}}%
}]{%
Ferranti2015blforcast}
\APACinsertmetastar {%
Ferranti2015blforcast}%
\begin{APACrefauthors}%
Ferranti, L.%
, Corti, S.%
\BCBL {}\ \BBA {} Janousek, M.%
\end{APACrefauthors}%
\unskip\
\newblock
\APACrefYearMonthDay{2015}{}{}.
\newblock
{\BBOQ}\APACrefatitle {{F}low-dependent verification of the {ECMWF} ensemble
  over the {E}uro-{A}tlantic sector} {{F}low-dependent verification of the
  {ECMWF} ensemble over the {E}uro-{A}tlantic sector}.{\BBCQ}
\newblock
\APACjournalVolNumPages{Quarterly Journal of the Royal Meteorological
  Society}{141}{688}{916-924}.
\newblock
\begin{APACrefURL}
  \url{https://rmets.onlinelibrary.wiley.com/doi/abs/10.1002/qj.2411}
  \end{APACrefURL}
\newblock
\begin{APACrefDOI} \doi{https://doi.org/10.1002/qj.2411} \end{APACrefDOI}
\PrintBackRefs{\CurrentBib}

\bibitem [\protect \citeauthoryear {%
Finkel%
, Webber%
, Gerber%
, Abbot%
\BCBL {}\ \BBA {} Weare%
}{%
Finkel%
\ \protect \BOthers {.}}{%
{\protect \APACyear {2023}}%
}]{%
Finkel2023}
\APACinsertmetastar {%
Finkel2023}%
\begin{APACrefauthors}%
Finkel, J.%
, Webber, R\BPBI J.%
, Gerber, E\BPBI P.%
, Abbot, D\BPBI S.%
\BCBL {}\ \BBA {} Weare, J.%
\end{APACrefauthors}%
\unskip\
\newblock
\APACrefYearMonthDay{2023}{}{}.
\newblock
{\BBOQ}\APACrefatitle {{D}ata-{D}riven {T}ransition {P}ath {A}nalysis {Y}ields
  a {S}tatistical {U}nderstanding of {S}udden {S}tratospheric {W}arming
  {E}vents in an {I}dealized {M}odel} {{D}ata-{D}riven {T}ransition {P}ath
  {A}nalysis {Y}ields a {S}tatistical {U}nderstanding of {S}udden
  {S}tratospheric {W}arming {E}vents in an {I}dealized {M}odel}.{\BBCQ}
\newblock
\APACjournalVolNumPages{Journal of the Atmospheric Sciences}{80}{2}{519 - 534}.
\newblock
\begin{APACrefURL}
  \url{https://journals.ametsoc.org/view/journals/atsc/80/2/JAS-D-21-0213.1.xml}
  \end{APACrefURL}
\newblock
\begin{APACrefDOI} \doi{https://doi.org/10.1175/JAS-D-21-0213.1}
  \end{APACrefDOI}
\PrintBackRefs{\CurrentBib}

\bibitem [\protect \citeauthoryear {%
González%
\ \protect \BOthers {.}}{%
González%
\ \protect \BOthers {.}}{%
{\protect \APACyear {2022}}%
}]{%
10062031}
\APACinsertmetastar {%
10062031}%
\begin{APACrefauthors}%
González, J\BPBI L.%
, Chapman, T.%
, Chen, K.%
, Nguyen, H.%
, Chambers, L.%
, Mostafa, S\BPBI A.%
\BDBL {}Yue, J.%
\end{APACrefauthors}%
\unskip\
\newblock
\APACrefYearMonthDay{2022}{}{}.
\newblock
{\BBOQ}\APACrefatitle {{A}tmospheric {G}ravity {W}ave {D}etection {U}sing
  {T}ransfer {L}earning {T}echniques} {{A}tmospheric {G}ravity {W}ave
  {D}etection {U}sing {T}ransfer {L}earning {T}echniques}.{\BBCQ}
\newblock
\BIn{} \APACrefbtitle {2022 {IEEE}/{ACM} {I}nternational {C}onference on {B}ig
  {D}ata {C}omputing, {A}pplications and {T}echnologies ({BDCAT})} {2022
  {IEEE}/{ACM} {I}nternational {C}onference on {B}ig {D}ata {C}omputing,
  {A}pplications and {T}echnologies ({BDCAT})}\ (\BPG~128-137).
\newblock
\begin{APACrefDOI} \doi{10.1109/BDCAT56447.2022.00023} \end{APACrefDOI}
\PrintBackRefs{\CurrentBib}

\bibitem [\protect \citeauthoryear {%
Goodfellow%
, Bengio%
\BCBL {}\ \BBA {} Courville%
}{%
Goodfellow%
\ \protect \BOthers {.}}{%
{\protect \APACyear {2016}}%
}]{%
Goodfellow-et-al-2016}
\APACinsertmetastar {%
Goodfellow-et-al-2016}%
\begin{APACrefauthors}%
Goodfellow, I.%
, Bengio, Y.%
\BCBL {}\ \BBA {} Courville, A.%
\end{APACrefauthors}%
\unskip\
\newblock
\APACrefYear{2016}.
\newblock
\APACrefbtitle {Deep Learning} {Deep learning}.
\newblock
\APACaddressPublisher{}{MIT Press}.
\newblock
\APACrefnote{\url{http://www.deeplearningbook.org}}
\PrintBackRefs{\CurrentBib}

\bibitem [\protect \citeauthoryear {%
Guo%
\ \protect \BOthers {.}}{%
Guo%
\ \protect \BOthers {.}}{%
{\protect \APACyear {2022}}%
}]{%
photonics9080582}
\APACinsertmetastar {%
photonics9080582}%
\begin{APACrefauthors}%
Guo, Y.%
, Wu, X.%
, Qing, C.%
, Su, C.%
, Yang, Q.%
\BCBL {}\ \BBA {} Wang, Z.%
\end{APACrefauthors}%
\unskip\
\newblock
\APACrefYearMonthDay{2022}{}{}.
\newblock
{\BBOQ}\APACrefatitle {{B}lind {R}estoration of {I}mages {D}istorted by
  {A}tmospheric {T}urbulence {B}ased on {D}eep {T}ransfer {L}earning} {{B}lind
  {R}estoration of {I}mages {D}istorted by {A}tmospheric {T}urbulence {B}ased
  on {D}eep {T}ransfer {L}earning}.{\BBCQ}
\newblock
\APACjournalVolNumPages{Photonics}{9}{8}{}.
\newblock
\begin{APACrefURL} \url{https://www.mdpi.com/2304-6732/9/8/582}
  \end{APACrefURL}
\newblock
\begin{APACrefDOI} \doi{10.3390/photonics9080582} \end{APACrefDOI}
\PrintBackRefs{\CurrentBib}

\bibitem [\protect \citeauthoryear {%
Ham%
, Kim%
\BCBL {}\ \BBA {} Luo%
}{%
Ham%
\ \protect \BOthers {.}}{%
{\protect \APACyear {2019}}%
}]{%
Ham2019}
\APACinsertmetastar {%
Ham2019}%
\begin{APACrefauthors}%
Ham, Y\BHBI G.%
, Kim, J\BHBI H.%
\BCBL {}\ \BBA {} Luo, J\BHBI J.%
\end{APACrefauthors}%
\unskip\
\newblock
\APACrefYearMonthDay{2019}{Sep}{01}.
\newblock
{\BBOQ}\APACrefatitle {{D}eep learning for multi-year {ENSO} forecasts} {{D}eep
  learning for multi-year {ENSO} forecasts}.{\BBCQ}
\newblock
\APACjournalVolNumPages{Nature}{573}{7775}{568-572}.
\newblock
\begin{APACrefURL} \url{https://doi.org/10.1038/s41586-019-1559-7}
  \end{APACrefURL}
\newblock
\begin{APACrefDOI} \doi{10.1038/s41586-019-1559-7} \end{APACrefDOI}
\PrintBackRefs{\CurrentBib}

\bibitem [\protect \citeauthoryear {%
Hersbach%
\ \protect \BOthers {.}}{%
Hersbach%
\ \protect \BOthers {.}}{%
{\protect \APACyear {2020}}%
}]{%
ERA5reanalysis2020}
\APACinsertmetastar {%
ERA5reanalysis2020}%
\begin{APACrefauthors}%
Hersbach, H.%
, Bell, B.%
, Berrisford, P.%
, Hirahara, S.%
, Horányi, A.%
, Muñoz-Sabater, J.%
\BDBL {}Thépaut, J\BHBI N.%
\end{APACrefauthors}%
\unskip\
\newblock
\APACrefYearMonthDay{2020}{}{}.
\newblock
{\BBOQ}\APACrefatitle {The ERA5 global reanalysis} {The era5 global
  reanalysis}.{\BBCQ}
\newblock
\APACjournalVolNumPages{Quarterly Journal of the Royal Meteorological
  Society}{146}{730}{1999-2049}.
\newblock
\begin{APACrefURL}
  \url{https://rmets.onlinelibrary.wiley.com/doi/abs/10.1002/qj.3803}
  \end{APACrefURL}
\newblock
\begin{APACrefDOI} \doi{https://doi.org/10.1002/qj.3803} \end{APACrefDOI}
\PrintBackRefs{\CurrentBib}

\bibitem [\protect \citeauthoryear {%
{Hoskins}%
, {James}%
\BCBL {}\ \BBA {} {White}%
}{%
{Hoskins}%
\ \protect \BOthers {.}}{%
{\protect \APACyear {1983}}%
}]{%
Hoskins1983}
\APACinsertmetastar {%
Hoskins1983}%
\begin{APACrefauthors}%
{Hoskins}, B\BPBI J.%
, {James}, I\BPBI N.%
\BCBL {}\ \BBA {} {White}, G\BPBI H.%
\end{APACrefauthors}%
\unskip\
\newblock
\APACrefYearMonthDay{1983}{{\APACmonth{07}}}{}.
\newblock
{\BBOQ}\APACrefatitle {{{T}he {S}hape, {P}ropagation and {M}ean-{F}low
  {I}nteraction of {L}arge-{S}cale {W}eather {S}ystems.}} {{{T}he {S}hape,
  {P}ropagation and {M}ean-{F}low {I}nteraction of {L}arge-{S}cale {W}eather
  {S}ystems.}}{\BBCQ}
\newblock
\APACjournalVolNumPages{Journal of Atmospheric Sciences}{40}{7}{1595-1612}.
\newblock
\begin{APACrefDOI} \doi{10.1175/1520-0469(1983)040<1595:TSPAMF>2.0.CO;2}
  \end{APACrefDOI}
\PrintBackRefs{\CurrentBib}

\bibitem [\protect \citeauthoryear {%
Hussain%
, Bird%
\BCBL {}\ \BBA {} Faria%
}{%
Hussain%
\ \protect \BOthers {.}}{%
{\protect \APACyear {2019}}%
}]{%
hussain2019study}
\APACinsertmetastar {%
hussain2019study}%
\begin{APACrefauthors}%
Hussain, M.%
, Bird, J\BPBI J.%
\BCBL {}\ \BBA {} Faria, D\BPBI R.%
\end{APACrefauthors}%
\unskip\
\newblock
\APACrefYearMonthDay{2019}{}{}.
\newblock
{\BBOQ}\APACrefatitle {{A} study on cnn transfer learning for image
  classification} {{A} study on cnn transfer learning for image
  classification}.{\BBCQ}
\newblock
\BIn{} \APACrefbtitle {{A}dvances in {C}omputational {I}ntelligence {S}ystems:
  {C}ontributions {P}resented at the 18th {UK} {W}orkshop on {C}omputational
  {I}ntelligence, {S}eptember 5-7, 2018, {N}ottingham, {UK}} {{A}dvances in
  {C}omputational {I}ntelligence {S}ystems: {C}ontributions {P}resented at the
  18th {UK} {W}orkshop on {C}omputational {I}ntelligence, {S}eptember 5-7,
  2018, {N}ottingham, {UK}}\ (\BPGS\ 191--202).
\PrintBackRefs{\CurrentBib}

\bibitem [\protect \citeauthoryear {%
Jacques-Dumas%
, Ragone%
, Borgnat%
, Abry%
\BCBL {}\ \BBA {} Bouchet%
}{%
Jacques-Dumas%
\ \protect \BOthers {.}}{%
{\protect \APACyear {2022}}%
}]{%
jacques2022deep}
\APACinsertmetastar {%
jacques2022deep}%
\begin{APACrefauthors}%
Jacques-Dumas, V.%
, Ragone, F.%
, Borgnat, P.%
, Abry, P.%
\BCBL {}\ \BBA {} Bouchet, F.%
\end{APACrefauthors}%
\unskip\
\newblock
\APACrefYearMonthDay{2022}{}{}.
\newblock
{\BBOQ}\APACrefatitle {{D}eep learning-based extreme heatwave forecast} {{D}eep
  learning-based extreme heatwave forecast}.{\BBCQ}
\newblock
\APACjournalVolNumPages{Frontiers in Climate}{4}{}{}.
\PrintBackRefs{\CurrentBib}

\bibitem [\protect \citeauthoryear {%
Johnson%
\ \BBA {} Khoshgoftaar%
}{%
Johnson%
\ \BBA {} Khoshgoftaar%
}{%
{\protect \APACyear {2019}}%
}]{%
johnson2019survey}
\APACinsertmetastar {%
johnson2019survey}%
\begin{APACrefauthors}%
Johnson, J\BPBI M.%
\BCBT {}\ \BBA {} Khoshgoftaar, T\BPBI M.%
\end{APACrefauthors}%
\unskip\
\newblock
\APACrefYearMonthDay{2019}{}{}.
\newblock
{\BBOQ}\APACrefatitle {{S}urvey on deep learning with class imbalance}
  {{S}urvey on deep learning with class imbalance}.{\BBCQ}
\newblock
\APACjournalVolNumPages{Journal of Big Data}{6}{1}{1--54}.
\PrintBackRefs{\CurrentBib}

\bibitem [\protect \citeauthoryear {%
Kautz%
\ \protect \BOthers {.}}{%
Kautz%
\ \protect \BOthers {.}}{%
{\protect \APACyear {2022}}%
}]{%
Woollings2022}
\APACinsertmetastar {%
Woollings2022}%
\begin{APACrefauthors}%
Kautz, L\BHBI A.%
, Martius, O.%
, Pfahl, S.%
, Pinto, J\BPBI G.%
, Ramos, A\BPBI M.%
, Sousa, P\BPBI M.%
\BCBL {}\ \BBA {} Woollings, T.%
\end{APACrefauthors}%
\unskip\
\newblock
\APACrefYearMonthDay{2022}{}{}.
\newblock
{\BBOQ}\APACrefatitle {{A}tmospheric blocking and weather extremes over the
  {E}uro-{A}tlantic sector -- a review} {{A}tmospheric blocking and weather
  extremes over the {E}uro-{A}tlantic sector -- a review}.{\BBCQ}
\newblock
\APACjournalVolNumPages{Weather and Climate Dynamics}{3}{1}{305--336}.
\newblock
\begin{APACrefURL} \url{https://wcd.copernicus.org/articles/3/305/2022/}
  \end{APACrefURL}
\newblock
\begin{APACrefDOI} \doi{10.5194/wcd-3-305-2022} \end{APACrefDOI}
\PrintBackRefs{\CurrentBib}

\bibitem [\protect \citeauthoryear {%
Labe%
\ \BBA {} Barnes%
}{%
Labe%
\ \BBA {} Barnes%
}{%
{\protect \APACyear {2021}}%
}]{%
explanableAIBarnes}
\APACinsertmetastar {%
explanableAIBarnes}%
\begin{APACrefauthors}%
Labe, Z\BPBI M.%
\BCBT {}\ \BBA {} Barnes, E\BPBI A.%
\end{APACrefauthors}%
\unskip\
\newblock
\APACrefYearMonthDay{2021}{}{}.
\newblock
{\BBOQ}\APACrefatitle {{D}etecting {C}limate {S}ignals {U}sing {E}xplainable
  {AI} {W}ith {S}ingle-{F}orcing {L}arge {E}nsembles} {{D}etecting {C}limate
  {S}ignals {U}sing {E}xplainable {AI} {W}ith {S}ingle-{F}orcing {L}arge
  {E}nsembles}.{\BBCQ}
\newblock
\APACjournalVolNumPages{Journal of Advances in Modeling Earth
  Systems}{13}{6}{e2021MS002464}.
\newblock
\begin{APACrefURL}
  \url{https://agupubs.onlinelibrary.wiley.com/doi/abs/10.1029/2021MS002464}
  \end{APACrefURL}
\newblock
\APACrefnote{e2021MS002464 2021MS002464}
\newblock
\begin{APACrefDOI} \doi{https://doi.org/10.1029/2021MS002464} \end{APACrefDOI}
\PrintBackRefs{\CurrentBib}

\bibitem [\protect \citeauthoryear {%
Linardatos%
, Papastefanopoulos%
\BCBL {}\ \BBA {} Kotsiantis%
}{%
Linardatos%
\ \protect \BOthers {.}}{%
{\protect \APACyear {2020}}%
}]{%
linardatos2020explainable}
\APACinsertmetastar {%
linardatos2020explainable}%
\begin{APACrefauthors}%
Linardatos, P.%
, Papastefanopoulos, V.%
\BCBL {}\ \BBA {} Kotsiantis, S.%
\end{APACrefauthors}%
\unskip\
\newblock
\APACrefYearMonthDay{2020}{}{}.
\newblock
{\BBOQ}\APACrefatitle {{E}xplainable ai: {A} review of machine learning
  interpretability methods} {{E}xplainable ai: {A} review of machine learning
  interpretability methods}.{\BBCQ}
\newblock
\APACjournalVolNumPages{Entropy}{23}{1}{18}.
\PrintBackRefs{\CurrentBib}

\bibitem [\protect \citeauthoryear {%
Lipovetsky%
\ \BBA {} Conklin%
}{%
Lipovetsky%
\ \BBA {} Conklin%
}{%
{\protect \APACyear {2001}}%
}]{%
ShapleyGametheory}
\APACinsertmetastar {%
ShapleyGametheory}%
\begin{APACrefauthors}%
Lipovetsky, S.%
\BCBT {}\ \BBA {} Conklin, M.%
\end{APACrefauthors}%
\unskip\
\newblock
\APACrefYearMonthDay{2001}{}{}.
\newblock
{\BBOQ}\APACrefatitle {{A}nalysis of regression in game theory approach}
  {{A}nalysis of regression in game theory approach}.{\BBCQ}
\newblock
\APACjournalVolNumPages{Applied Stochastic Models in Business and
  Industry}{17}{4}{319-330}.
\newblock
\begin{APACrefURL}
  \url{https://onlinelibrary.wiley.com/doi/abs/10.1002/asmb.446}
  \end{APACrefURL}
\newblock
\begin{APACrefDOI} \doi{https://doi.org/10.1002/asmb.446} \end{APACrefDOI}
\PrintBackRefs{\CurrentBib}

\bibitem [\protect \citeauthoryear {%
Liu%
\ \protect \BOthers {.}}{%
Liu%
\ \protect \BOthers {.}}{%
{\protect \APACyear {2016}}%
}]{%
liu2016application}
\APACinsertmetastar {%
liu2016application}%
\begin{APACrefauthors}%
Liu, Y.%
, Racah, E.%
, Prabhat%
, Correa, J.%
, Khosrowshahi, A.%
, Lavers, D.%
\BDBL {}Collins, W.%
\end{APACrefauthors}%
\unskip\
\newblock
\APACrefYearMonthDay{2016}{}{}.
\newblock
\APACrefbtitle {{A}pplication of {D}eep {C}onvolutional {N}eural {N}etworks for
  {D}etecting {E}xtreme {W}eather in {C}limate {D}atasets.} {{A}pplication of
  {D}eep {C}onvolutional {N}eural {N}etworks for {D}etecting {E}xtreme
  {W}eather in {C}limate {D}atasets.}
\PrintBackRefs{\CurrentBib}

\bibitem [\protect \citeauthoryear {%
Lucarini%
\ \BBA {} Gritsun%
}{%
Lucarini%
\ \BBA {} Gritsun%
}{%
{\protect \APACyear {2020}}%
}]{%
lucarini2020new}
\APACinsertmetastar {%
lucarini2020new}%
\begin{APACrefauthors}%
Lucarini, V.%
\BCBT {}\ \BBA {} Gritsun, A.%
\end{APACrefauthors}%
\unskip\
\newblock
\APACrefYearMonthDay{2020}{}{}.
\newblock
{\BBOQ}\APACrefatitle {{A} new mathematical framework for atmospheric blocking
  events} {{A} new mathematical framework for atmospheric blocking
  events}.{\BBCQ}
\newblock
\APACjournalVolNumPages{Climate Dynamics}{54}{1-2}{575--598}.
\PrintBackRefs{\CurrentBib}

\bibitem [\protect \citeauthoryear {%
Lundberg%
\ \BBA {} Lee%
}{%
Lundberg%
\ \BBA {} Lee%
}{%
{\protect \APACyear {2017}}%
}]{%
lundberg2017unified}
\APACinsertmetastar {%
lundberg2017unified}%
\begin{APACrefauthors}%
Lundberg, S\BPBI M.%
\BCBT {}\ \BBA {} Lee, S\BHBI I.%
\end{APACrefauthors}%
\unskip\
\newblock
\APACrefYearMonthDay{2017}{}{}.
\newblock
{\BBOQ}\APACrefatitle {{A} unified approach to interpreting model predictions}
  {{A} unified approach to interpreting model predictions}.{\BBCQ}
\newblock
\APACjournalVolNumPages{Advances in neural information processing
  systems}{30}{}{}.
\PrintBackRefs{\CurrentBib}

\bibitem [\protect \citeauthoryear {%
Lupo%
}{%
Lupo%
}{%
{\protect \APACyear {2021}}%
}]{%
Lupo2021}
\APACinsertmetastar {%
Lupo2021}%
\begin{APACrefauthors}%
Lupo, A\BPBI R.%
\end{APACrefauthors}%
\unskip\
\newblock
\APACrefYearMonthDay{2021}{}{}.
\newblock
{\BBOQ}\APACrefatitle {{A}tmospheric blocking events: a review} {{A}tmospheric
  blocking events: a review}.{\BBCQ}
\newblock
\APACjournalVolNumPages{Annals of the New York Academy of
  Sciences}{1504}{1}{5-24}.
\newblock
\begin{APACrefURL}
  \url{https://nyaspubs.onlinelibrary.wiley.com/doi/abs/10.1111/nyas.14557}
  \end{APACrefURL}
\newblock
\begin{APACrefDOI} \doi{https://doi.org/10.1111/nyas.14557} \end{APACrefDOI}
\PrintBackRefs{\CurrentBib}

\bibitem [\protect \citeauthoryear {%
Lupo%
\ \protect \BOthers {.}}{%
Lupo%
\ \protect \BOthers {.}}{%
{\protect \APACyear {2012}}%
}]{%
lupo2012dynamic}
\APACinsertmetastar {%
lupo2012dynamic}%
\begin{APACrefauthors}%
Lupo, A\BPBI R.%
, Mokhov, I\BPBI I.%
, Akperov, M\BPBI G.%
, Chernokulsky, A\BPBI V.%
, Athar, H.%
\BCBL {}\ \BOthersPeriod {.}\end{APACrefauthors}%
\unskip\
\newblock
\APACrefYearMonthDay{2012}{}{}.
\newblock
{\BBOQ}\APACrefatitle {{A} dynamic analysis of the role of the planetary-and
  synoptic-scale in the summer of 2010 blocking episodes over the {E}uropean
  part of {R}ussia} {{A} dynamic analysis of the role of the planetary-and
  synoptic-scale in the summer of 2010 blocking episodes over the {E}uropean
  part of {R}ussia}.{\BBCQ}
\newblock
\APACjournalVolNumPages{Advances in Meteorology}{2012}{}{}.
\PrintBackRefs{\CurrentBib}

\bibitem [\protect \citeauthoryear {%
Malmgren-Hansen%
, Nielsen%
, Laparra%
\BCBL {}\ \BBA {} Valls%
}{%
Malmgren-Hansen%
\ \protect \BOthers {.}}{%
{\protect \APACyear {2018}}%
}]{%
8518097}
\APACinsertmetastar {%
8518097}%
\begin{APACrefauthors}%
Malmgren-Hansen, D.%
, Nielsen, A\BPBI A.%
, Laparra, V.%
\BCBL {}\ \BBA {} Valls, G\BPBI C.%
\end{APACrefauthors}%
\unskip\
\newblock
\APACrefYearMonthDay{2018}{}{}.
\newblock
{\BBOQ}\APACrefatitle {{T}ransfer {L}earning with {C}onvolutional {N}etworks
  for {A}tmospheric {P}arameter {R}etrieval} {{T}ransfer {L}earning with
  {C}onvolutional {N}etworks for {A}tmospheric {P}arameter {R}etrieval}.{\BBCQ}
\newblock
\BIn{} \APACrefbtitle {{IGARSS} 2018 - 2018 {IEEE} {I}nternational {G}eoscience
  and {R}emote {S}ensing {S}ymposium} {{IGARSS} 2018 - 2018 {IEEE}
  {I}nternational {G}eoscience and {R}emote {S}ensing {S}ymposium}\
  (\BPG~2111-2114).
\newblock
\begin{APACrefDOI} \doi{10.1109/IGARSS.2018.8518097} \end{APACrefDOI}
\PrintBackRefs{\CurrentBib}

\bibitem [\protect \citeauthoryear {%
Marshall%
\ \BBA {} Molteni%
}{%
Marshall%
\ \BBA {} Molteni%
}{%
{\protect \APACyear {1993}}%
}]{%
marshall1993toward}
\APACinsertmetastar {%
marshall1993toward}%
\begin{APACrefauthors}%
Marshall, J.%
\BCBT {}\ \BBA {} Molteni, F.%
\end{APACrefauthors}%
\unskip\
\newblock
\APACrefYearMonthDay{1993}{}{}.
\newblock
{\BBOQ}\APACrefatitle {{T}oward a dynamical understanding of planetary-scale
  flow regimes} {{T}oward a dynamical understanding of planetary-scale flow
  regimes}.{\BBCQ}
\newblock
\APACjournalVolNumPages{Journal of the atmospheric
  sciences}{50}{12}{1792--1818}.
\PrintBackRefs{\CurrentBib}

\bibitem [\protect \citeauthoryear {%
Matsueda%
}{%
Matsueda%
}{%
{\protect \APACyear {2009}}%
}]{%
Mediumrange2009}
\APACinsertmetastar {%
Mediumrange2009}%
\begin{APACrefauthors}%
Matsueda, M.%
\end{APACrefauthors}%
\unskip\
\newblock
\APACrefYearMonthDay{2009}{}{}.
\newblock
{\BBOQ}\APACrefatitle {{B}locking {P}redictability in {O}perational
  {M}edium-{R}ange {E}nsemble {F}orecasts} {{B}locking {P}redictability in
  {O}perational {M}edium-{R}ange {E}nsemble {F}orecasts}.{\BBCQ}
\newblock
\APACjournalVolNumPages{SOLA}{5}{}{113-116}.
\newblock
\begin{APACrefDOI} \doi{10.2151/sola.2009-029} \end{APACrefDOI}
\PrintBackRefs{\CurrentBib}

\bibitem [\protect \citeauthoryear {%
McWilliams%
}{%
McWilliams%
}{%
{\protect \APACyear {1980}}%
}]{%
MCWILLIAMS198043}
\APACinsertmetastar {%
MCWILLIAMS198043}%
\begin{APACrefauthors}%
McWilliams, J\BPBI C.%
\end{APACrefauthors}%
\unskip\
\newblock
\APACrefYearMonthDay{1980}{}{}.
\newblock
{\BBOQ}\APACrefatitle {{A}n application of equivalent modons to atmospheric
  blocking} {{A}n application of equivalent modons to atmospheric
  blocking}.{\BBCQ}
\newblock
\APACjournalVolNumPages{Dynamics of Atmospheres and Oceans}{5}{1}{43-66}.
\newblock
\begin{APACrefURL}
  \url{https://www.sciencedirect.com/science/article/pii/037702658090010X}
  \end{APACrefURL}
\newblock
\begin{APACrefDOI} \doi{https://doi.org/10.1016/0377-0265(80)90010-X}
  \end{APACrefDOI}
\PrintBackRefs{\CurrentBib}

\bibitem [\protect \citeauthoryear {%
Michelangeli%
\ \BBA {} Vautard%
}{%
Michelangeli%
\ \BBA {} Vautard%
}{%
{\protect \APACyear {1998}}%
}]{%
michelangeli1998dynamics}
\APACinsertmetastar {%
michelangeli1998dynamics}%
\begin{APACrefauthors}%
Michelangeli, P\BHBI A.%
\BCBT {}\ \BBA {} Vautard, R.%
\end{APACrefauthors}%
\unskip\
\newblock
\APACrefYearMonthDay{1998}{}{}.
\newblock
{\BBOQ}\APACrefatitle {{T}he dynamics of {E}uro-{A}tlantic blocking onsets}
  {{T}he dynamics of {E}uro-{A}tlantic blocking onsets}.{\BBCQ}
\newblock
\APACjournalVolNumPages{Quarterly Journal of the Royal Meteorological
  Society}{124}{548}{1045--1070}.
\PrintBackRefs{\CurrentBib}

\bibitem [\protect \citeauthoryear {%
Miloshevich%
, Cozian%
, Abry%
, Borgnat%
\BCBL {}\ \BBA {} Bouchet%
}{%
Miloshevich%
\ \protect \BOthers {.}}{%
{\protect \APACyear {2023}}%
}]{%
Bouchet2023}
\APACinsertmetastar {%
Bouchet2023}%
\begin{APACrefauthors}%
Miloshevich, G.%
, Cozian, B.%
, Abry, P.%
, Borgnat, P.%
\BCBL {}\ \BBA {} Bouchet, F.%
\end{APACrefauthors}%
\unskip\
\newblock
\APACrefYearMonthDay{2023}{Apr}{}.
\newblock
{\BBOQ}\APACrefatitle {{P}robabilistic forecasts of extreme heatwaves using
  convolutional neural networks in a regime of lack of data} {{P}robabilistic
  forecasts of extreme heatwaves using convolutional neural networks in a
  regime of lack of data}.{\BBCQ}
\newblock
\APACjournalVolNumPages{Phys. Rev. Fluids}{8}{}{040501}.
\newblock
\begin{APACrefURL}
  \url{https://link.aps.org/doi/10.1103/PhysRevFluids.8.040501}
  \end{APACrefURL}
\newblock
\begin{APACrefDOI} \doi{10.1103/PhysRevFluids.8.040501} \end{APACrefDOI}
\PrintBackRefs{\CurrentBib}

\bibitem [\protect \citeauthoryear {%
Mu%
, Ma%
, Yuan%
\BCBL {}\ \BBA {} Xu%
}{%
Mu%
\ \protect \BOthers {.}}{%
{\protect \APACyear {2020}}%
}]{%
Mu2020}
\APACinsertmetastar {%
Mu2020}%
\begin{APACrefauthors}%
Mu, B.%
, Ma, S.%
, Yuan, S.%
\BCBL {}\ \BBA {} Xu, H.%
\end{APACrefauthors}%
\unskip\
\newblock
\APACrefYearMonthDay{2020}{}{}.
\newblock
{\BBOQ}\APACrefatitle {Applying Convolutional LSTM Network to Predict El Niño
  Events: Transfer Learning from The Data of Dynamical Model and Observation}
  {Applying convolutional lstm network to predict el niño events: Transfer
  learning from the data of dynamical model and observation}.{\BBCQ}
\newblock
\BIn{} \APACrefbtitle {2020 IEEE 10th International Conference on Electronics
  Information and Emergency Communication (ICEIEC)} {2020 ieee 10th
  international conference on electronics information and emergency
  communication (iceiec)}\ (\BPG~215-219).
\newblock
\begin{APACrefDOI} \doi{10.1109/ICEIEC49280.2020.9152317} \end{APACrefDOI}
\PrintBackRefs{\CurrentBib}

\bibitem [\protect \citeauthoryear {%
Mullen%
}{%
Mullen%
}{%
{\protect \APACyear {1987}}%
}]{%
mullen1987transient}
\APACinsertmetastar {%
mullen1987transient}%
\begin{APACrefauthors}%
Mullen, S\BPBI L.%
\end{APACrefauthors}%
\unskip\
\newblock
\APACrefYearMonthDay{1987}{}{}.
\newblock
{\BBOQ}\APACrefatitle {{T}ransient eddy forcing of blocking flows} {{T}ransient
  eddy forcing of blocking flows}.{\BBCQ}
\newblock
\APACjournalVolNumPages{Journal of the Atmospheric Sciences}{44}{1}{3--22}.
\PrintBackRefs{\CurrentBib}

\bibitem [\protect \citeauthoryear {%
Pelly%
\ \BBA {} Hoskins%
}{%
Pelly%
\ \BBA {} Hoskins%
}{%
{\protect \APACyear {2003}}%
}]{%
pelly2003new}
\APACinsertmetastar {%
pelly2003new}%
\begin{APACrefauthors}%
Pelly, J\BPBI L.%
\BCBT {}\ \BBA {} Hoskins, B\BPBI J.%
\end{APACrefauthors}%
\unskip\
\newblock
\APACrefYearMonthDay{2003}{}{}.
\newblock
{\BBOQ}\APACrefatitle {{A} new perspective on blocking} {{A} new perspective on
  blocking}.{\BBCQ}
\newblock
\APACjournalVolNumPages{Journal of the atmospheric sciences}{60}{5}{743--755}.
\PrintBackRefs{\CurrentBib}

\bibitem [\protect \citeauthoryear {%
Rampal%
\ \protect \BOthers {.}}{%
Rampal%
\ \protect \BOthers {.}}{%
{\protect \APACyear {2022}}%
}]{%
RAMPAL2022100525}
\APACinsertmetastar {%
RAMPAL2022100525}%
\begin{APACrefauthors}%
Rampal, N.%
, Gibson, P\BPBI B.%
, Sood, A.%
, Stuart, S.%
, Fauchereau, N\BPBI C.%
, Brandolino, C.%
\BDBL {}Meyers, T.%
\end{APACrefauthors}%
\unskip\
\newblock
\APACrefYearMonthDay{2022}{}{}.
\newblock
{\BBOQ}\APACrefatitle {{H}igh-resolution downscaling with interpretable deep
  learning: {R}ainfall extremes over {N}ew {Z}ealand} {{H}igh-resolution
  downscaling with interpretable deep learning: {R}ainfall extremes over {N}ew
  {Z}ealand}.{\BBCQ}
\newblock
\APACjournalVolNumPages{Weather and Climate Extremes}{38}{}{100525}.
\newblock
\begin{APACrefURL}
  \url{https://www.sciencedirect.com/science/article/pii/S2212094722001049}
  \end{APACrefURL}
\newblock
\begin{APACrefDOI} \doi{https://doi.org/10.1016/j.wace.2022.100525}
  \end{APACrefDOI}
\PrintBackRefs{\CurrentBib}

\bibitem [\protect \citeauthoryear {%
Rasp%
\ \BBA {} Thuerey%
}{%
Rasp%
\ \BBA {} Thuerey%
}{%
{\protect \APACyear {2021}}%
}]{%
Prediction2020}
\APACinsertmetastar {%
Prediction2020}%
\begin{APACrefauthors}%
Rasp, S.%
\BCBT {}\ \BBA {} Thuerey, N.%
\end{APACrefauthors}%
\unskip\
\newblock
\APACrefYearMonthDay{2021}{}{}.
\newblock
{\BBOQ}\APACrefatitle {{D}ata-{D}riven {M}edium-{R}ange {W}eather {P}rediction
  {W}ith a {R}esnet {P}retrained on {C}limate {S}imulations: {A} {N}ew {M}odel
  for {W}eather{B}ench} {{D}ata-{D}riven {M}edium-{R}ange {W}eather
  {P}rediction {W}ith a {R}esnet {P}retrained on {C}limate {S}imulations: {A}
  {N}ew {M}odel for {W}eather{B}ench}.{\BBCQ}
\newblock
\APACjournalVolNumPages{Journal of Advances in Modeling Earth
  Systems}{13}{2}{e2020MS002405}.
\newblock
\begin{APACrefURL}
  \url{https://agupubs.onlinelibrary.wiley.com/doi/abs/10.1029/2020MS002405}
  \end{APACrefURL}
\newblock
\APACrefnote{e2020MS002405 2020MS002405}
\newblock
\begin{APACrefDOI} \doi{https://doi.org/10.1029/2020MS002405} \end{APACrefDOI}
\PrintBackRefs{\CurrentBib}

\bibitem [\protect \citeauthoryear {%
Rex%
}{%
Rex%
}{%
{\protect \APACyear {1950}}%
}]{%
Rex1950}
\APACinsertmetastar {%
Rex1950}%
\begin{APACrefauthors}%
Rex, D\BPBI F.%
\end{APACrefauthors}%
\unskip\
\newblock
\APACrefYearMonthDay{1950}{}{}.
\newblock
{\BBOQ}\APACrefatitle {{B}locking {A}ction in the {M}iddle {T}roposphere and
  its {E}ffect upon {R}egional {C}limate} {{B}locking {A}ction in the {M}iddle
  {T}roposphere and its {E}ffect upon {R}egional {C}limate}.{\BBCQ}
\newblock
\APACjournalVolNumPages{Tellus}{2}{3}{196-211}.
\newblock
\begin{APACrefURL}
  \url{https://onlinelibrary.wiley.com/doi/abs/10.1111/j.2153-3490.1950.tb00331.x}
  \end{APACrefURL}
\newblock
\begin{APACrefDOI} \doi{https://doi.org/10.1111/j.2153-3490.1950.tb00331.x}
  \end{APACrefDOI}
\PrintBackRefs{\CurrentBib}

\bibitem [\protect \citeauthoryear {%
Rudy%
\ \BBA {} Sapsis%
}{%
Rudy%
\ \BBA {} Sapsis%
}{%
{\protect \APACyear {2023}}%
}]{%
rudy2023output}
\APACinsertmetastar {%
rudy2023output}%
\begin{APACrefauthors}%
Rudy, S\BPBI H.%
\BCBT {}\ \BBA {} Sapsis, T\BPBI P.%
\end{APACrefauthors}%
\unskip\
\newblock
\APACrefYearMonthDay{2023}{}{}.
\newblock
{\BBOQ}\APACrefatitle {{O}utput-weighted and relative entropy loss functions
  for deep learning precursors of extreme events} {{O}utput-weighted and
  relative entropy loss functions for deep learning precursors of extreme
  events}.{\BBCQ}
\newblock
\APACjournalVolNumPages{Physica D: Nonlinear Phenomena}{443}{}{133570}.
\PrintBackRefs{\CurrentBib}

\bibitem [\protect \citeauthoryear {%
Scaife%
, Woollings%
, Knight%
, Martin%
\BCBL {}\ \BBA {} Hinton%
}{%
Scaife%
\ \protect \BOthers {.}}{%
{\protect \APACyear {2010}}%
}]{%
scaife2010atmospheric}
\APACinsertmetastar {%
scaife2010atmospheric}%
\begin{APACrefauthors}%
Scaife, A\BPBI A.%
, Woollings, T.%
, Knight, J.%
, Martin, G.%
\BCBL {}\ \BBA {} Hinton, T.%
\end{APACrefauthors}%
\unskip\
\newblock
\APACrefYearMonthDay{2010}{}{}.
\newblock
{\BBOQ}\APACrefatitle {{A}tmospheric blocking and mean biases in climate
  models} {{A}tmospheric blocking and mean biases in climate models}.{\BBCQ}
\newblock
\APACjournalVolNumPages{Journal of Climate}{23}{23}{6143--6152}.
\PrintBackRefs{\CurrentBib}

\bibitem [\protect \citeauthoryear {%
Shrikumar%
, Greenside%
\BCBL {}\ \BBA {} Kundaje%
}{%
Shrikumar%
\ \protect \BOthers {.}}{%
{\protect \APACyear {2017}}%
}]{%
shrikumar2017learning}
\APACinsertmetastar {%
shrikumar2017learning}%
\begin{APACrefauthors}%
Shrikumar, A.%
, Greenside, P.%
\BCBL {}\ \BBA {} Kundaje, A.%
\end{APACrefauthors}%
\unskip\
\newblock
\APACrefYearMonthDay{2017}{}{}.
\newblock
{\BBOQ}\APACrefatitle {{L}earning important features through propagating
  activation differences} {{L}earning important features through propagating
  activation differences}.{\BBCQ}
\newblock
\BIn{} \APACrefbtitle {{I}nternational conference on machine learning}
  {{I}nternational conference on machine learning}\ (\BPGS\ 3145--3153).
\PrintBackRefs{\CurrentBib}

\bibitem [\protect \citeauthoryear {%
Shutts%
}{%
Shutts%
}{%
{\protect \APACyear {1983}}%
}]{%
shutts1983propagation}
\APACinsertmetastar {%
shutts1983propagation}%
\begin{APACrefauthors}%
Shutts, G.%
\end{APACrefauthors}%
\unskip\
\newblock
\APACrefYearMonthDay{1983}{}{}.
\newblock
{\BBOQ}\APACrefatitle {{T}he propagation of eddies in diffluent jetstreams:
  {E}ddy vorticity forcing of ‘blocking’flow fields} {{T}he propagation of
  eddies in diffluent jetstreams: {E}ddy vorticity forcing of
  ‘blocking’flow fields}.{\BBCQ}
\newblock
\APACjournalVolNumPages{Quarterly Journal of the Royal Meteorological
  Society}{109}{462}{737--761}.
\PrintBackRefs{\CurrentBib}

\bibitem [\protect \citeauthoryear {%
Silva%
, Keller%
\BCBL {}\ \BBA {} Hardin%
}{%
Silva%
\ \protect \BOthers {.}}{%
{\protect \APACyear {2022}}%
}]{%
silva2022using}
\APACinsertmetastar {%
silva2022using}%
\begin{APACrefauthors}%
Silva, S\BPBI J.%
, Keller, C\BPBI A.%
\BCBL {}\ \BBA {} Hardin, J.%
\end{APACrefauthors}%
\unskip\
\newblock
\APACrefYearMonthDay{2022}{}{}.
\newblock
{\BBOQ}\APACrefatitle {{U}sing an explainable machine learning approach to
  characterize {E}arth {S}ystem model errors: {A}pplication of {SHAP} analysis
  to modeling lightning flash occurrence} {{U}sing an explainable machine
  learning approach to characterize {E}arth {S}ystem model errors:
  {A}pplication of {SHAP} analysis to modeling lightning flash
  occurrence}.{\BBCQ}
\newblock
\APACjournalVolNumPages{Journal of Advances in Modeling Earth
  Systems}{14}{4}{e2021MS002881}.
\PrintBackRefs{\CurrentBib}

\bibitem [\protect \citeauthoryear {%
Subel%
, Chattopadhyay%
, Guan%
\BCBL {}\ \BBA {} Hassanzadeh%
}{%
Subel%
\ \protect \BOthers {.}}{%
{\protect \APACyear {2021}}%
}]{%
subel2021data}
\APACinsertmetastar {%
subel2021data}%
\begin{APACrefauthors}%
Subel, A.%
, Chattopadhyay, A.%
, Guan, Y.%
\BCBL {}\ \BBA {} Hassanzadeh, P.%
\end{APACrefauthors}%
\unskip\
\newblock
\APACrefYearMonthDay{2021}{}{}.
\newblock
{\BBOQ}\APACrefatitle {{D}ata-driven subgrid-scale modeling of forced {B}urgers
  turbulence using deep learning with generalization to higher {R}eynolds
  numbers via transfer learning} {{D}ata-driven subgrid-scale modeling of
  forced {B}urgers turbulence using deep learning with generalization to higher
  {R}eynolds numbers via transfer learning}.{\BBCQ}
\newblock
\APACjournalVolNumPages{Physics of Fluids}{33}{3}{}.
\PrintBackRefs{\CurrentBib}

\bibitem [\protect \citeauthoryear {%
Talo%
, Baloglu%
, Y{\i}ld{\i}r{\i}m%
\BCBL {}\ \BBA {} Acharya%
}{%
Talo%
\ \protect \BOthers {.}}{%
{\protect \APACyear {2019}}%
}]{%
talo2019application}
\APACinsertmetastar {%
talo2019application}%
\begin{APACrefauthors}%
Talo, M.%
, Baloglu, U\BPBI B.%
, Y{\i}ld{\i}r{\i}m, {\"O}.%
\BCBL {}\ \BBA {} Acharya, U\BPBI R.%
\end{APACrefauthors}%
\unskip\
\newblock
\APACrefYearMonthDay{2019}{}{}.
\newblock
{\BBOQ}\APACrefatitle {{A}pplication of deep transfer learning for automated
  brain abnormality classification using {MR} images} {{A}pplication of deep
  transfer learning for automated brain abnormality classification using {MR}
  images}.{\BBCQ}
\newblock
\APACjournalVolNumPages{Cognitive Systems Research}{54}{}{176--188}.
\PrintBackRefs{\CurrentBib}

\bibitem [\protect \citeauthoryear {%
Tibaldi%
\ \BBA {} Molteni%
}{%
Tibaldi%
\ \BBA {} Molteni%
}{%
{\protect \APACyear {1990}}%
}]{%
tibaldi1990operational}
\APACinsertmetastar {%
tibaldi1990operational}%
\begin{APACrefauthors}%
Tibaldi, S.%
\BCBT {}\ \BBA {} Molteni, F.%
\end{APACrefauthors}%
\unskip\
\newblock
\APACrefYearMonthDay{1990}{}{}.
\newblock
{\BBOQ}\APACrefatitle {{O}n the operational predictability of blocking} {{O}n
  the operational predictability of blocking}.{\BBCQ}
\newblock
\APACjournalVolNumPages{Tellus A}{42}{3}{343--365}.
\PrintBackRefs{\CurrentBib}

\bibitem [\protect \citeauthoryear {%
Vitart%
\ \protect \BOthers {.}}{%
Vitart%
\ \protect \BOthers {.}}{%
{\protect \APACyear {2017}}%
}]{%
Vitart2017subseasonal}
\APACinsertmetastar {%
Vitart2017subseasonal}%
\begin{APACrefauthors}%
Vitart, F.%
, Ardilouze, C.%
, Bonet, A.%
, Brookshaw, A.%
, Chen, M.%
, Codorean, C.%
\BDBL {}Zhang, L.%
\end{APACrefauthors}%
\unskip\
\newblock
\APACrefYearMonthDay{2017}{}{}.
\newblock
{\BBOQ}\APACrefatitle {{T}he {S}ubseasonal to {S}easonal ({S}2{S}) {P}rediction
  {P}roject {D}atabase} {{T}he {S}ubseasonal to {S}easonal ({S}2{S})
  {P}rediction {P}roject {D}atabase}.{\BBCQ}
\newblock
\APACjournalVolNumPages{Bulletin of the American Meteorological
  Society}{98}{1}{163 - 173}.
\newblock
\begin{APACrefURL}
  \url{https://journals.ametsoc.org/view/journals/bams/98/1/bams-d-16-0017.1.xml}
  \end{APACrefURL}
\newblock
\begin{APACrefDOI} \doi{10.1175/BAMS-D-16-0017.1} \end{APACrefDOI}
\PrintBackRefs{\CurrentBib}

\bibitem [\protect \citeauthoryear {%
Woollings%
\ \protect \BOthers {.}}{%
Woollings%
\ \protect \BOthers {.}}{%
{\protect \APACyear {2018}}%
}]{%
Woollings2018}
\APACinsertmetastar {%
Woollings2018}%
\begin{APACrefauthors}%
Woollings, T.%
, Barriopedro, D.%
, Methven, J.%
, Son, S\BHBI W.%
, Martius, O.%
, Harvey, B.%
\BDBL {}Seneviratne, S.%
\end{APACrefauthors}%
\unskip\
\newblock
\APACrefYearMonthDay{2018}{Sep}{01}.
\newblock
{\BBOQ}\APACrefatitle {{B}locking and its {R}esponse to {C}limate {C}hange}
  {{B}locking and its {R}esponse to {C}limate {C}hange}.{\BBCQ}
\newblock
\APACjournalVolNumPages{Current Climate Change Reports}{}{}{}.
\PrintBackRefs{\CurrentBib}

\bibitem [\protect \citeauthoryear {%
Yang%
\ \protect \BOthers {.}}{%
Yang%
\ \protect \BOthers {.}}{%
{\protect \APACyear {2021}}%
}]{%
yang_influence_2021}
\APACinsertmetastar {%
yang_influence_2021}%
\begin{APACrefauthors}%
Yang, M.%
, Luo, D.%
, Li, C.%
, Yao, Y.%
, Li, X.%
\BCBL {}\ \BBA {} Chen, X.%
\end{APACrefauthors}%
\unskip\
\newblock
\APACrefYearMonthDay{2021}{}{}.
\newblock
{\BBOQ}\APACrefatitle {{I}nfluence of {{A}tmospheric} {{B}locking} on {{S}torm}
  {{T}rack} {{A}ctivity} {{O}ver} the {{N}orth} {{P}acific} {{D}uring}
  {{B}oreal} {{W}inter}} {{I}nfluence of {{A}tmospheric} {{B}locking} on
  {{S}torm} {{T}rack} {{A}ctivity} {{O}ver} the {{N}orth} {{P}acific}
  {{D}uring} {{B}oreal} {{W}inter}}.{\BBCQ}
\newblock
\APACjournalVolNumPages{Geophysical Research Letters}{48}{17}{e2021GL093863}.
\newblock
\begin{APACrefURL}
  [{2023-08-03}]\url{https://onlinelibrary.wiley.com/doi/abs/10.1029/2021GL093863}
  \end{APACrefURL}
\newblock
\APACrefnote{\_eprint:
  https://onlinelibrary.wiley.com/doi/pdf/10.1029/2021GL093863}
\newblock
\begin{APACrefDOI} \doi{10.1029/2021GL093863} \end{APACrefDOI}
\PrintBackRefs{\CurrentBib}

\bibitem [\protect \citeauthoryear {%
Yosinski%
, Clune%
, Bengio%
\BCBL {}\ \BBA {} Lipson%
}{%
Yosinski%
\ \protect \BOthers {.}}{%
{\protect \APACyear {2014}}%
}]{%
yosinski2014transferable}
\APACinsertmetastar {%
yosinski2014transferable}%
\begin{APACrefauthors}%
Yosinski, J.%
, Clune, J.%
, Bengio, Y.%
\BCBL {}\ \BBA {} Lipson, H.%
\end{APACrefauthors}%
\unskip\
\newblock
\APACrefYearMonthDay{2014}{}{}.
\newblock
{\BBOQ}\APACrefatitle {{H}ow transferable are features in deep neural
  networks?} {{H}ow transferable are features in deep neural networks?}{\BBCQ}
\newblock
\APACjournalVolNumPages{Advances in neural information processing
  systems}{27}{}{}.
\PrintBackRefs{\CurrentBib}

\bibitem [\protect \citeauthoryear {%
Zappa%
, Masato%
, Shaffrey%
, Woollings%
\BCBL {}\ \BBA {} Hodges%
}{%
Zappa%
\ \protect \BOthers {.}}{%
{\protect \APACyear {2014}}%
{\protect \APACexlab {{\protect \BCnt {1}}}}}]{%
zappa2014linking}
\APACinsertmetastar {%
zappa2014linking}%
\begin{APACrefauthors}%
Zappa, G.%
, Masato, G.%
, Shaffrey, L.%
, Woollings, T.%
\BCBL {}\ \BBA {} Hodges, K.%
\end{APACrefauthors}%
\unskip\
\newblock
\APACrefYearMonthDay{2014{\protect \BCnt {1}}}{}{}.
\newblock
{\BBOQ}\APACrefatitle {{L}inking {N}orthern {H}emisphere blocking and storm
  track biases in the {CMIP}5 climate models} {{L}inking {N}orthern
  {H}emisphere blocking and storm track biases in the {CMIP}5 climate
  models}.{\BBCQ}
\newblock
\APACjournalVolNumPages{Geophysical Research Letters}{41}{1}{135--139}.
\PrintBackRefs{\CurrentBib}

\bibitem [\protect \citeauthoryear {%
Zappa%
, Masato%
, Shaffrey%
, Woollings%
\BCBL {}\ \BBA {} Hodges%
}{%
Zappa%
\ \protect \BOthers {.}}{%
{\protect \APACyear {2014}}%
{\protect \APACexlab {{\protect \BCnt {2}}}}}]{%
zappa_linking_2014}
\APACinsertmetastar {%
zappa_linking_2014}%
\begin{APACrefauthors}%
Zappa, G.%
, Masato, G.%
, Shaffrey, L.%
, Woollings, T.%
\BCBL {}\ \BBA {} Hodges, K.%
\end{APACrefauthors}%
\unskip\
\newblock
\APACrefYearMonthDay{2014{\protect \BCnt {2}}}{}{}.
\newblock
{\BBOQ}\APACrefatitle {{L}inking {{N}orthern} {{H}emisphere} blocking and storm
  track biases in the {{CMIP}5} climate models} {{L}inking {{N}orthern}
  {{H}emisphere} blocking and storm track biases in the {{CMIP}5} climate
  models}.{\BBCQ}
\newblock
\APACjournalVolNumPages{Geophysical Research Letters}{41}{1}{135--139}.
\newblock
\begin{APACrefURL}
  [{2023-08-02}]\url{https://onlinelibrary.wiley.com/doi/abs/10.1002/2013GL058480}
  \end{APACrefURL}
\newblock
\APACrefnote{\_eprint:
  https://onlinelibrary.wiley.com/doi/pdf/10.1002/2013GL058480}
\newblock
\begin{APACrefDOI} \doi{10.1002/2013GL058480} \end{APACrefDOI}
\PrintBackRefs{\CurrentBib}

\end{thebibliography}

\end{document}